\documentclass[11pt,reprint,showkeys, amsmath,amssymb, aps, nofootinbib, floatfix]{revtex4-2}
\DeclareUnicodeCharacter{2212}{\textendash}
\usepackage{graphicx,subfigure}
\usepackage[export]{adjustbox}
\usepackage{dcolumn}
\usepackage{bm}
\usepackage{physics}
\usepackage{xcolor}
\usepackage{slashed,cancel}
\usepackage{multirow}
\usepackage{float}
\usepackage{capt-of}
\usepackage{comment}
\usepackage{mathrsfs}
\usepackage{enumitem}

\usepackage[colorlinks=true,linkcolor=blue,urlcolor=blue,citecolor=blue]{hyperref}
\usepackage{orcidlink}

\usepackage{tikz}
\usepackage{lmodern}
\usetikzlibrary{
  arrows.meta,
  backgrounds,
  calc,
  decorations.pathreplacing,
  positioning,
  shapes.geometric
}

\allowdisplaybreaks
\begin{document}
\preprint{APS/123-QED}

\title{Probing SMEFT Operators through $t\bar{t}t\bar{t}$ Production  with Hyper-Graph Neural Networks at the LHC}

\author{Amir Subba\,\orcidlink{https://orcid.org/0000-0002-0126-1419}}
\email{amirsubba@ustc.edu.cn}
\affiliation{Wilczek Quantum Center, Shanghai Institute for Advanced Studies, Shanghai 201315, China}
\affiliation{University of Science and Technology of China, Hefei 230026, China }

\author{Sanmay Ganguly\,\orcidlink{https://orcid.org/0000-0003-1285-9261}}
\email{sanmay@iitk.ac.in}
\affiliation{Department of Physics, Indian Institute of Technology Kanpur, Uttar Pradesh 208016, India}

\date{\today}

\begin{abstract}
The production of four top quarks ($t\bar{t}t\bar{t}$) at hadron colliders constitutes one of the rarest Standard Model (SM) processes accessible at the
Large Hadron Collider (LHC) and provides a uniquely sensitive probe of the top-quark Yukawa coupling, of its possible $CP$ structure, and of new
physics encoded in four-heavy-fermion operators within the Standard Model Effective Field Theory (SMEFT) framework. We present a phenomenological
study of $t\bar{t}t\bar{t}$ production in proton-proton collisions at $\sqrt{s} = 13$~TeV, using a Hyper-Graph Neural Network (H-GNN) to
discriminate multilepton signal events from the dominant SM backgrounds viz. $t\bar{t}W$, $t\bar{t}Z$, $t\bar{t}H$ and $t\bar{t}VV$ processes.
In the H-GNN architecture each event is represented as a hypergraph whose nodes correspond to reconstructed jets and leptons and whose hyperedges
encode higher-order correlations among arbitrary subsets of these objects, allowing the network to learn the many-body kinematic structures that
characterize the $t\bar{t}t\bar{t}$ final state. Combining same-sign di-lepton, tri-lepton, and four-lepton channels following a CMS-like event
selection, the H-GNN attains an area under the ROC curve of $0.97$ for the $t\bar{t}t\bar{t}$ signal and yields a statistical significance of $Z=6.8$ at $\mathcal{L}=140$~fb$^{-1}$. We exploit the improved signal extraction to derive one-parameter $95\%$ confidence level limits on the Wilson coefficients of the dimension-six operators $\mathcal{O}_{\Phi u}$, $\mathcal{O}^{(1)}_{tt}$, $\mathcal{O}^{(1)}_{qq}$, $\mathcal{O}^{(1)}_{qt}$, and $\mathcal{O}^{(8)}_{qt}$. We project the expected sensitivity at the HL-LHC integrated luminosities of $1000~\mathrm{fb}^{-1}$ and $3000~\mathrm{fb}^{-1}$ with $50\%$ uncertainty on the background estimation. Finally, the sensitivity reach of H-GNN is compared to the SPANet and Particle Transformer baseline. 
\end{abstract}

\maketitle

\section{Introduction}
\label{sec:intro}
The top quark, by virtue of its mass close to the electroweak symmetry
breaking scale~\cite{ParticleDataGroup:2024cfk}, occupies a privileged position within the Standard Model
(SM) and in essentially every viable extension of it. Its Yukawa coupling
to the Higgs boson is of order unity~\cite{ParticleDataGroup:2024cfk}, making it a natural portal between
the visible sector and any new physics that couples preferentially to the
electroweak-symmetry-breaking sector. Moreover, its decay width is large
enough that the top quark decays before hadronization~\cite{Grossman:2008qh}, so that its spin and
chirality information is transmitted faithfully to its decay products and
becomes experimentally accessible. These features make top-quark
production a privileged laboratory both for precision tests of the SM and
for indirect searches for physics beyond it.

\begin{figure*}[!htb]
    \centering
    \includegraphics[width=0.32\linewidth,height=4.5cm]{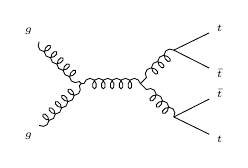}
    \includegraphics[width=0.32\linewidth,height=4.5cm]{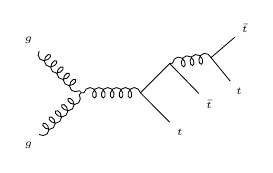}
    \includegraphics[width=0.32\linewidth,height=4.5cm]{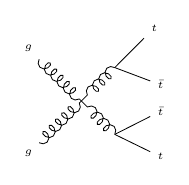}
    \includegraphics[width=0.32\linewidth,height=4.5cm]{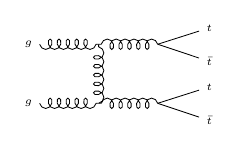}
    \includegraphics[width=0.32\linewidth,height=4.5cm]{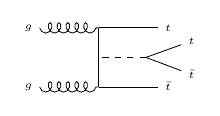}
    \includegraphics[width=0.32\linewidth,height=4.5cm]{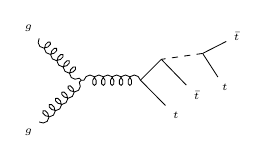}
    \caption{Schematic representation of some of the leading order Feynman diagrams for the production of four top quarks in the Standard Model at hadron collider.}
    \label{fig:feynman}
\end{figure*}

Among the variety of top-quark processes accessible at the LHC, the
simultaneous production of two top-antitop pairs, $pp \to t\bar{t}t\bar{t}$,
is of particular theoretical and phenomenological interest. At leading
order in the SM, this process is dominated by gluon-initiated QCD
amplitudes, with subleading but non-negligible contributions from
electroweak diagrams~\cite{Bevilacqua:2012em,Alwall:2014hca,Maltoni:2015ena}.
A subset of these electroweak amplitudes involves an internal Higgs boson
exchanged between two top--antitop lines, see Fig.~\ref{fig:feynman}; the corresponding contribution scales as $y_t^4$ and the interference with the dominant QCD amplitude as $y_t^2 g_s^4$, with $y_t$ the top-Yukawa coupling and $g_s$ the strong coupling. This quartic dependence on $y_t$ is to be contrasted with the at-most quadratic sensitivity carried by $t\bar{t}H$ associated production and the still milder dependence of inclusive $t\bar{t}$ differential observables, and is the origin of the strong leverage that $t\bar{t}t\bar{t}$ events offer in constraining anomalous top--Higgs interactions and possible $CP$-violating components of the top Yukawa~\cite{Martini:2021uey}.

The cross section for $t\bar{t}t\bar{t}$ production at $\sqrt{s} = 13$~TeV
has been computed at next-to-leading order (NLO) in QCD including
electroweak corrections to be
$\sigma_{t\bar{t}t\bar{t}}^{\mathrm{NLO}} = 12 \pm 2.4~\mathrm{fb}$, with
the uncertainty dominated by missing higher-order
contributions estimated from variations of the renormalization and
factorization
scales~\cite{Bevilacqua:2012em,Alwall:2014hca,Maltoni:2015ena,Frederix:2017wme,Jezo:2021smh}. This prediction places $t\bar{t}t\bar{t}$ among the rarest processes accessible at the LHC, smaller than inclusive $t\bar{t}$ by roughly five orders of magnitude. The observation of the process has therefore become an experimental milestone in its own right: the ATLAS collaboration~\cite{ATLAS:2023ajo} has reported the observation of $t\bar{t}t\bar{t}$ in multilepton final states with a measured cross section of $\sigma_{t\bar{t}t\bar{t}} = 22.5^{+6.6}_{-5.5}~\mathrm{fb}$ at a significance of $6.1$ standard deviations, and the
CMS collaboration~\cite{CMS:2023ftu} has reported a measurement of
$\sigma_{t\bar{t}t\bar{t}} = 17 \pm 5~\mathrm{fb}$ at $5.6$ standard
deviations, both at $\sqrt{s} = 13$~TeV and with an
integrated luminosity of $140~\mathrm{fb}^{-1}$. The measured central
values lie above, but are statistically compatible with, the SM prediction.

Beyond the intrinsic interest of $t\bar{t}t\bar{t}$ as the SM
measurement, this process is also a powerful probe of physics beyond the
SM. We adopt the SMEFT framework, in which deviations from the SM are
parameterized in terms of higher-dimensional gauge-invariant operators
built from SM fields~\cite{Buchmuller:1985jz}. Truncating
the expansion at dimension six and setting the cut-off scale to
$\Lambda = 1~\mathrm{TeV}$, the operator $\mathcal{O}_{u\Phi}$~\cite{Grzadkowski:2010es} modifies
the top-Yukawa coupling, including its $CP$ structure, while the
four-heavy-fermion operators $\mathcal{O}^{(1)}_{tt}$,
$\mathcal{O}^{(1)}_{qq}$, $\mathcal{O}^{(1)}_{qt}$, and
$\mathcal{O}^{(8)}_{qt}$~\cite{DHondt:2018cww}, generate contact interactions among third generation quarks that interfere coherently with the SM amplitude and
shape the high-energy tails of the $t\bar{t}t\bar{t}$ differential
distributions. The strong dependence of the $t\bar{t}t\bar{t}$ rate on the
top-Yukawa is already used by ATLAS~\cite{ATLAS:2023ajo} to set the observed bound $|\kappa_t| < 2.2$ at $95\%$ confidence level, while
current one-parameter limits on the Wilson coefficients (WCs) of the four
heavy-fermion operators lie in the range
$\mathcal{O}(1)$--$\mathcal{O}(10)~\mathrm{TeV}^{-2}$ and are tabulated in
Section~\ref{sec:smeft}.

The extraction of these constraints is, however, limited by experimental
challenges intrinsic to the process. The small SM cross section, the high
jet and $b$-jet multiplicity, and the presence of irreducible backgrounds
from $t\bar{t}W$, $t\bar{t}Z$, $t\bar{t}H$, $t\bar{t}VV$, and single-top
associated production yield signal regions in which the
signal-to-background ratio is typically $\mathcal{O}(10^{-2})$ even after
stringent multilepton, $H_T$ and $b$-tagging requirements. Furthermore, the
combinatorial ambiguity inherent in associating up to twelve final-state
partons to four parent top quarks makes the construction of fully
reconstructed kinematic observables difficult, and limits the
discriminating power of conventional cut-based and low-dimensional
multivariate analyses. These features motivate the use of modern machine
learning techniques capable of exploiting the full event information in a
data-driven and high-dimensional manner.

Among such techniques, Graph Neural Networks (GNNs)~\cite{gnnScar} have emerged in recent years as a particularly powerful class of architectures for
collider physics applications. 
A hard-scattering event admits a natural
graph representation in which reconstructed objects viz. jets, leptons, missing transverse momentum are nodes and their pairwise relations are encoded as edges. This representation is permutation invariant, respects the variable cardinality of collider final states, and aligns naturally with
the locality of QCD radiation patterns. However, the physically relevant
correlations in a $t\bar{t}t\bar{t}$ event are intrinsically many-body in
nature. The invariant mass of a $b\ell\nu$ triplet, the angular pattern
of two $b$-jets and two same-sign leptons, or the imbalance between the
hadronic and leptonic transverse momentum sums are observables that
involve three or more reconstructed objects simultaneously. These features
cannot be captured by pairwise edges alone without iterated message
passing of substantial depth. Hypergraphs, in which a single
\emph{hyperedge} connects an arbitrary subset of nodes, generalize graphs
in a manner that makes such many-body correlations directly representable
at the architectural level. The Hyper-Graph Neural Network (H-GNN)~\cite{Feng:2018hgnn} 
based architecture, we
introduce in this work exploits this property to learn the higher-order
kinematic structures that characterize the $t\bar{t}t\bar{t}$ final state
without the inductive bias toward two-body interactions implicit in
standard GNNs.

Although hypergraph methods are by now well established in computer
vision and recommendation systems, their adoption in high-energy
physics is recent and so far confined to reconstruction tasks rather
than event-level classification. The HGPflow algorithm of
Ref.~\cite{DiBello:2022iwf} introduced hypergraph prediction networks
for particle-flow reconstruction inside individual jets, exploiting
the natural correspondence between hyperedges and the one-to-many
association of calorimeter cells and tracks to particles. 
In a complementary direction, the HyPER architecture of Ref.~\cite{Birch-Sykes:2024gij}
applies hypergraph representation learning to the reconstruction of
heavy, short lived particles from their hadronic decay products,
demonstrating that the hypergraph inductive bias delivers competitive
assignment accuracy at substantially reduced parameter cost relative
to attention-based assignment networks such as SPANet~\cite{Shmakov:2021qdz}. 
Physics guided hypergraph based transformer methods have been proposed for 
future LHC runs (HL-LHC) in the Ref.~\cite{Rakib:2026ero} to model top quark pair production.
As an application to jet substructure methods, information geometry based observables
on higher order structures using hypergraphs, has been recently proposed in Ref.~\cite{Bal:2026pzw}.

In this article, we develop a dedicated H-GNN tagger for $t\bar{t}t\bar{t}$
events at the LHC and use it as the basis for an SMEFT analysis at
$\sqrt{s} = 13$~TeV. Following the CMS event-selection
strategy~\cite{CMS:2023ftu}, we combine the same-sign dilepton, trilepton,
and four-lepton signal regions and feed the resulting events to the H-GNN.
We benchmark the H-GNN against two state-of-the-art alternative
architectures, the Particle Transformer (ParT)~\cite{Qu:2022mxj} and SPANet, trained and evaluated
under identical conditions, and find that the H-GNN delivers the highest
classification performance. H-GNN achieves an area under the ROC curve of $0.97$
for $t\bar{t}t\bar{t}$ versus the dominant $t\bar{t}V$ ($V \in W, Z$), $t\bar{t}H$ and $t\bar{t}VV$ backgrounds. At an integrated luminosity of
$\mathcal{L} = 140~\mathrm{fb}^{-1}$ the H-GNN achieves an optimal statistical
significance of $Z=6.8$
compared to $Z=6.1$ by the ATLAS~\cite{ATLAS:2023ajo} analysis, evaluated under identical event selection. The improved signal extraction
translates directly into tighter constraints on the SMEFT parameter
space: we report one-parameter $95\%$ confidence-level limits on
the WCs $C_{u \Phi}/\Lambda^2$, $C^{(1)}_{tt}/\Lambda^2$,
$C^{(1)}_{qq}/\Lambda^2$, $C^{(1)}_{qt}/\Lambda^2$, and
$C^{(8)}_{qt}/\Lambda^2$ and also project the expected reach of the
high-luminosity LHC at $\mathcal{L} = 1000~\mathrm{fb}^{-1}$ and
$3000~\mathrm{fb}^{-1}$.

The remainder of this paper is organized as follows. In
Section~\ref{sec:smeft} we describe the SMEFT framework used to
parameterize possible deviations from the SM in $t\bar{t}t\bar{t}$
production, introducing the relevant dimension-six operators and the
matching to the anomalous top--Higgs couplings. Section~\ref{sec:event}
details the signal and background simulation chain, the choice of
electroweak input scheme, and the multilepton event selection.
Section~\ref{sec:hgnn} introduces the Hyper-Graph Neural Network
architecture, motivates the choice of input features, and describes the
training procedure; the tagger performance is compared with Particle
Transformer and SPANet baselines on the same data set. In
Section~\ref{sec:result} we present a binned likelihood analysis of the
tagged events and derive constraints on the Wilson coefficients at present
and projected luminosities. We conclude in Section~\ref{sec:conclusion}
with a summary of our results and an outlook on possible extensions of
this work. Appendix~\ref{app:networks} collects additional comparisons
between the three network architectures considered.
\section{SMEFT production of four top events}
\label{sec:smeft}
In the SM at leading order (LO), \(t\bar{t}t\bar{t}\) arises dominantly from gluon‑initiated QCD amplitudes, however a subclass of amplitudes with an intermediate virtual Higgs boson coupling to a top‑quark pair exists, see Fig.~\ref{fig:feynman}. The contribution of these EW amplitudes are proportional to \(y_t^4\) and mixed terms proportional to \(y_t^2 g_s^4\). This quartic dependence on the top Yukawa coupling contrasts with processes such as \(t\bar{t}H\), where the dependence is approximately \(y_t^2\), or differential \(t\bar{t}\) observables sensitive only through weak corrections, thereby enhancing the leverage of the $t\bar{t}t\bar{t}$ process in constraining anomalous top-Yukawa. 

In the current work, we parameterize the deviations from the SM in terms of higher dimensional operators constructed from the SM fields and satisfying the gauge symmetry of the SM. Such framework is usually know as SM effective field theory (SMEFT) with an assumption that the EW symmetry breaking proceeds linearly. Within the SMEFT, the effective Lagrangian is written as~\cite{Buchmuller:1985jz}
\begin{equation}
    \label{eq:smeft}
    \mathcal{L} = \mathcal{L}_{\mathrm{SM}} + \sum_i \frac{C_i^{(d)}}{\Lambda^{d-4}} \mathscr{O}_i^{(d)},
\end{equation}
where $C_i^{(d)}$ corresponds to the set of Wilson coefficients associated with set of operators $\mathscr{O}_i^{(d)}$ in dimension $d$ and $\Lambda$ is the characteristic cut-off scale, with $\Lambda=1$ TeV set for current analysis. We truncate the above Eq.~\eqref{eq:smeft} at $d=6$, where $d=5$ is ignored considering the baryon number conservation.

At $d=6$, there exist one operator affecting the top-Yukawa coupling~\cite{Grzadkowski:2010es},
\begin{equation}
    \mathscr{O}_{u \Phi} = (\Phi^\dagger\Phi)\,(q_pu_r\widetilde{\Phi}),
\end{equation}
where $\Phi$ is the SU(2)$_{\mathrm{L}}$ doublet which in the unitary gauge is defined as 
\begin{align}
    \Phi = \frac{1}{\sqrt{2}}
    \begin{pmatrix}
        0 \\ v+H
    \end{pmatrix},
\end{align}
where $v$ is the vacuum expectation value and $H$ is the quanta of Higgs field. The complex conjugate of the Higgs filed is defined as $\widetilde{\Phi} = \epsilon_{jk}(\Phi^k)^\star$, where $\epsilon_{jk}$ is totally anti-symmetric with $\epsilon_{12}=+1$. The deviation in the top-Yukawa from the SM are also parameterized in terms of two anomalous parameters as~\cite{Martini:2021uey}
\begin{equation}
    \label{eq:ltth}
    \mathcal{L}_{t\bar{t}H} \supset -\frac{m_t}{v}\bar{t}\left(a_t + i b_t\gamma_5\right)tH,
\end{equation}
where $a_t$ is the CP-even term and $b_t$ is the CP-odd parameter. The essential motivation for the CP-violating term in effective top-Higgs interaction is based on the observation that during the electroweak phase transition in the early universe, $\mathscr{L}_{t\bar{t}H}$ can help produce the baryon number asymmetry. The parameters $a_t$ and $b_t$ can be connected to the real and imaginary WCs of $\mathscr{O}_{\Phi u}$ as~\cite{Martini:2021uey}
\begin{align}
    \label{eq:invert}
    a_t &= 1-\frac{v}{\sqrt{2}m_t}\frac{v^2}{\Lambda^2}\,\mathrm{Re}\,\left[C_{u\Phi }\right],\nonumber \\
    b_t &= -\frac{v}{\sqrt{2}m_t} \frac{v^2}{\Lambda^2}\,\mathrm{Im}\,\left[C_{u\Phi }\right].
\end{align}
The amount of CP-violation of the top-Yukawa can be quantified as~\cite{Martini:2021uey}
\begin{equation}
    f_{\mathrm{CP}} = \frac{|b_t|^2}{|a_t|^2+|b_t|^2}\,\mathrm{sign}\left(\frac{b_t}{a_t}\right).
\end{equation}
Recent measurement by ATLAS~\cite{ATLAS:2023ajo} considered $a_t = \kappa_t \cos(\alpha)$ and $b_t = \kappa_t \sin(\alpha)$, with $\kappa_t$ being the top-Yukawa coupling modifier and $\alpha$ the CP-mixing angle. The observed $95\%$ confidence level (CL) limits on $|\kappa_t|$ by ATLAS~\cite{ATLAS:2023ajo} with $t\bar{t}t\bar{t}$ process is $|\kappa_t| < 2.2$ with an integrated luminosity $\mathcal{L} = 140$ fb$^{-1}$ at $\sqrt{s}=13$ TeV.

Apart from the top-Yukawa, the $pp\to t\bar{t}t\bar{t}$ process is sensitive to four heavy-flavor fermion operators~\cite{DHondt:2018cww}
\begin{align}
    \label{eq:heavy}
    \mathscr{O}_{tt}^{(1)} &= \Big(\bar{t}_R \gamma_\mu t_R\Big)\Big(\bar{t}_R\gamma^\mu t_R\Big),\nonumber\\
    \mathscr{O}_{qt}^{(1)} &= \Big(\bar{q}_L\gamma_\mu q_L\Big)\Big(\bar{t}_R\gamma^\mu t_R\Big),\nonumber\\
    \mathscr{O}_{qt}^{(8)} &= \Big(\bar{q}_L\gamma_\mu T^a q_L\Big)\Big(\bar{t}_R\gamma^\mu T^at_R\Big),\nonumber\\
    \mathscr{O}_{qq}^{(1)} &= \Big(\bar{q}_L\gamma_\mu q_L\Big)\Big(\bar{q}_L \gamma^\mu q_L\Big),
\end{align}
where $q$ represent the left-handed $SU(2)_L$ doublet if third generation quarks, $t_R$ is the right handed top quark and $T^a$ are the $SU(3)_C$ generators. The current best experimental limits on the WCs of the above four contact operator are listed in Table~\ref{tab:atlassmeft}.
\begin{table}[!h]
    \centering
    \renewcommand{\arraystretch}{1.3}
    \caption{\label{tab:atlassmeft} Observed one parameter $95\%$ CL intervals on WCs of four contact operators by ATLAS~\cite{ATLAS:2023ajo} with $t\bar{t}t\bar{t}$ process at $\sqrt{s}=13$ TeV and $\mathcal{L}=140$ fb$^{-1}$.}
    \begin{tabular*}{0.49\textwidth}{@{\extracolsep{\fill}}cc@{}}\hline
         Wilson Coefficients & Observed $C_i/\Lambda^2$ [TeV$^{-2}$]  \\
         \hline
        $C_{tt}^{(1)}$  & $[-1.7,+1.9]$ \\
        $C_{qq}^{(1)}$  & $[-3.5,+4.1]$ \\
        $C_{qt}^{(1)}$ & $[-3.5,+3.0]$ \\
        $C_{qt}^{(8)}$  & $[-6.2,+6.9]$ \\
        \hline
    \end{tabular*}
\end{table}
\section{Collider Simulation}
\label{sec:event}

All simulations are performed for proton--proton collisions at a center-of-mass energy of $\sqrt{s} = 13~\mathrm{TeV}$, targeting multi-lepton final states arising from four top quark production. Hard-scattering matrix elements are generated using \textsc{MadGraph5\_aMC@NLO}~\cite{Alwall:2011uj} at the leading order for both the SM background processes and the anomalous benchmark scenarios. The parton-level events are subsequently interfaced with \textsc{Pythia8}~\cite{Bierlich:2022pfr} for parton showering and hadronization, ensuring a consistent treatment of QCD radiation and final-state fragmentation. Detector acceptance and resolution effects are emulated with \textsc{Delphes}~\cite{deFavereau:2013fsa} using a CMS-like parametrization, which encodes realistic object reconstruction efficiencies and momentum smearing representative of a general-purpose LHC detector.

The SM backgrounds are chosen to comprehensively cover all processes capable of producing multi-lepton final states with $b$-tagged jets. The dominant contributions arise from $t\bar{t}V$ associated production (with $V \in \{W, Z, \gamma\}$), and $t\bar{t}H$, all of which share the same high jet-multiplicity and $b$-jet topology as the signal.
We have also considered the $t\bar{t}VV$ process, which emulates the signal topology.
Single-top associated production channels viz. $tW$, and $tWZ$ are included to account for single-top contributions to the multi-lepton yield. Diboson ($VV$) and triboson ($VVV$) production are incorporated to model prompt lepton backgrounds from purely electroweak sources. Finally, rare processes involving three top quarks in association with a light jet or a $W$ boson are included to account for high-multiplicity tails.

The electroweak sector is parameterized within the $\alpha$-scheme, following the conventions established at LEP~\cite{Biekotter:2023xle}. In this scheme, the triplet $\{\hat{\alpha}_\mathrm{EW},\, \hat{G}_F,\, \hat{m}_Z\}$ serves as the fundamental input from which all SM gauge couplings are derived self-consistently
\begin{align}
\hat{e} &= \sqrt{4\pi \hat{\alpha}_{\mathrm{EW}}}, \quad
\hat{g}_1 = \frac{\hat{e}}{\cos\hat{\theta}}, \quad
\hat{g}_2 = \frac{\hat{e}}{\sin\hat{\theta}}, \nonumber\\
v &= \frac{1}{2^{1/4}\sqrt{\hat{G}_F}},\quad \hat{m}_W^2 = \hat{m}_Z^2\cos^2\hat{\theta}, \nonumber\\
\sin^2\hat{\theta} &= \frac{1}{2}\left[1-\sqrt{1-\frac{4\pi\hat{\alpha}_{\mathrm{EW}}}{\sqrt{2}\,\hat{G}_F\,\hat{m}_Z^2}}\right],
\end{align}
where $v$ is the Higgs vacuum expectation value and $\hat{\theta}$ is the weak mixing angle. The numerical values of the SM input parameters are set as
\begin{align}
    m_Z &= 91.19~\mathrm{GeV},\quad \alpha_{\mathrm{EW}}^{-1} = 132.5,\nonumber \\ G_F&=1.1664\times 10^{-5}~\mathrm{GeV}^{-2},\nonumber \\ m_H&=125~\mathrm{GeV},\quad m_t=173~\mathrm{GeV},\quad \alpha_S &= 0.13.
\end{align}

Jets are clustered using the anti-$k_T$ algorithm~\cite{Cacciari:2008gp} with a radius parameter $R = 0.4$, and $b$-jets are identified via a standard $b$-tagging criteria implemented in \textsc{Delphes}. Jets with $p_T^j \ge 25.0$~GeV and $|\eta_j| \le 2.5$ are selected. Leptons (electrons and muons) with $p_T^\ell \ge 10.0$~GeV and $|\eta_\ell| \le 2.5$ are selected. The leptons are also required to satisfy isolation criteria to suppress contributions from non-prompt leptons originating from heavy-flavor decays or hadron misidentification. The isolation variable is defined as~\cite{deFavereau:2013fsa}:
\begin{equation}
I(\ell) = \displaystyle\sum^{\substack{\Delta R < R,\; p_T^i > p_T^{\mathrm{min}}}}_{i\neq \ell } p_T^i/p_T^{\ell},
\label{eq:isolation}
\end{equation}
where $p_T^{\ell}$ is the transverse momentum of the lepton under consideration, and the sum runs over all other particles within a cone of radius $\Delta R$ around the lepton axis satisfying $p_T^i > p_T^\mathrm{min}$. We adopt $p_T^{\mathrm{min}} = 0.5~\mathrm{GeV}$ and $\Delta R = 0.5$ uniformly for both electrons and muons. For electron, we choose $I(e)^{\mathrm{max}} = 0.12$, while for muon it is set to $0.25$.

Following the CMS multi-lepton analysis strategy of Ref.~\cite{CMS:2023ftu}, events are categorized into three exclusive signal regions (SR) according to the lepton-jet multiplicity and charge configuration.

\textbf{SR-$2\ell$:} Events are required to contain exactly two isolated leptons with same charge. This same-sign requirement strongly suppresses the dominant Drell--Yan and $t\bar{t}$ backgrounds. Further kinematic requirements of $N_j \geq 4$, $N_b \geq 2$, and $H_T > 280~\mathrm{GeV}$ are imposed, where $H_T$ denotes the scalar sum of jet transverse momenta. To enhance sensitivity specifically to high-multiplicity top-quark final states, at least one of the additional conditions $N_j \geq 6$ or $N_b \geq 3$ must be satisfied.

\textbf{SR-$3\ell$:} Events must contain exactly three isolated leptons, with requirements of $N_j \geq 3$, $N_b \geq 2$, and $H_T > 200~\mathrm{GeV}$. To suppress backgrounds from $Z$-boson production in association with jets or top quarks, events are vetoed if any opposite-sign same-flavor (OSSF) lepton pair is found with invariant mass satisfying $|m_{\ell\ell} - m_Z| < 15~\mathrm{GeV}$, effectively removing the $Z$-boson mass window.

\textbf{SR-$4\ell$:} Events must contain exactly four isolated leptons with a total lepton charge summing to zero, selecting configurations consistent with the production of two opposite-sign pairs of top quarks. Additional requirements of $N_j \geq 2$ and $N_b \geq 1$ are imposed. As in SR-$3\ell$, a $Z$-mass veto on any OSSF lepton pair is applied to reduce contamination from $ZZ$ and $t\bar{t}Z$ backgrounds in which both $Z$ bosons decay leptonically.

The corresponding cut-flow table showing the  signal and background yields after each event selection criterion for three different channels are shown in Table~\ref{tab:cut-flow}. The selection column $N_{\mathrm{Cut}}^{i\ell},i\in\{2,3,4\}$ represents the initial multiplicity cuts in each channels as discussed above. The selection efficiency are obtained for inclusive decay of resonances using \textsc{Pythia} with reconstructed objects at \textsc{Delphes}.

\begin{table*}[!htb]
    \centering
    \caption{Cut-flow table showing the signal and background yield after each selection criteria in three different channel. Events are generated using \textsc{MG5} at $\sqrt{s}=13$ TeV followed by inclusive decay with \textsc{Pythia8} and detector effects are implemented via \textsc{Delphes}.}
    \label{tab:cut-flow}
    \renewcommand{\arraystretch}{1.4}
    \begin{tabular}{c|c|c|c|c|c|c|c|c}
        \hline
        & \multicolumn{3}{c|}{Same-Sign $2\ell$}
        & \multicolumn{3}{c|}{$3\ell$}
        & \multicolumn{2}{c}{$4\ell$}
        \\
        \cline{2-4}
        \cline{5-7}
        \cline{8-9}
        \raisebox{1.3ex}{Process}
        & $N_{\mathrm{Cut}}^{(2\ell)}$
        & $H_T > 280$ GeV
        & $N_j \ge 6$ or $N_b \ge 3$
        & $N_{\mathrm{Cut}}^{(3\ell)}$
        & $Z$ Veto
        & $H_T > 200$ GeV
        & $N_{\mathrm{Cut}}^{(4\ell)}$
        & $Z$ Veto
        \\
        \hline
        $t\bar{t}t\bar{t}$& $0.0164$ & $0.0162$ & $0.0146$ & $0.0052$ & $0.0043$ & $0.0043$ & $0.0003$ & $0.0002$\\
        $t\bar{t}Z$ & $0.0013$ & $0.0010$ & $0.0005$ & $0.0024$ & $0.0002$  & $0.0002$ &$0.0005$ & $0.0001$ \\
        $t\bar{t}W$ & $0.0034$& $0.0024$ & $0.0007$  & $0.0004$ &$0.0003$ & $0.0003$ & $\sim 0.0$ & $\sim 0.0$ \\
        $t\bar{t}H$ &$0.0019$ &$0.0016$ &$0.0008$ &$0.0008$ & $0.0005$  & $0.0005$  &$0.0001$ & $0.0001$ \\
        $t\bar{t}VV$&$0.0042$&$0.0037$&$0.0019$&$0.0022$&$0.0012$&$0.0011$&$0.0004$&$0.0001$\\
         $t\bar{t}\gamma$ &$0.0002$&$0.0001$ &$\sim 0.0$&$0.0003$ & $0.0002$ & $0.0002$ & $0.0$ & $0.0$\\
        $tX$ &$\sim 0.0$&$0.0$ &$ 0.0$&$ 0.0$ & $0.0$ & $0.0$ & $0.0$ & $0.0$\\
        $t\bar{t}$ & $\sim 0.0$ & $0.0$ & $0.0$ & $0.0$ & $0.0$ &$0.0$ &$0.0$ &$0.0$ \\
         $VV(V)$ &$0.0$ &$0.0$ & $0.0$ & $0.0$& $0.0$& $0.0$& $0.0$&$0.0$ \\
        \hline
    \end{tabular}
\end{table*}

The combined effect of the lepton-jet multiplicity requirements, charge selection, $b$-jet multiplicity thresholds, and $Z$-mass vetoes is a substantial suppression of backgrounds with limited genuine multi-lepton and $b$-jet content. Processes such as diboson ($VV$), triboson ($VVV$),  single-top associated production ($tW$, $tWZ$) and multi-top rare processes ($t\bar{t}$, $t\bar{t}\gamma$) contribute negligibly to the selected event yields after these requirements. Consequently, the four dominant irreducible backgrounds retained for further analysis are $t\bar{t}W$, $t\bar{t}Z$, $t\bar{t}H$ and $t\bar{t}VV$, all of which share genuine high-lepton-multiplicity and high-$b$-jet-multiplicity topologies with the signal and cannot be efficiently rejected without significant signal loss. Furthermore, to increase the number of events for analysis,  we consider the leptonic decay of the processes considered above. Events from all three signal regions, along with these three background processes and the signal benchmark points, are combined and provided as input to the Hyper-Graph neural network (H-GNN) described in the following section.

\section{Hyper-Graph Neural Network}
\label{sec:hgnn}

In this section we describe the H-GNN
classifier developed to discriminate $t\bar{t}t\bar{t}$ events from the
dominant Standard Model backgrounds in the combined multilepton signal
region, viz. Same-Sign $2\ell$, $3\ell$ and $4\ell$, defined in Section~\ref{sec:event}. We first motivate the choice of a hypergraph representation of the reconstructed event
(Sec.~\ref{sec:hgnn:motivation}), specify the construction of nodes and
hyperedges (Sec.~\ref{sec:hgnn:representation}), describe the network
architecture (Sec.~\ref{sec:hgnn:architecture}) and the training
procedure (Sec.~\ref{sec:hgnn:training}), and conclude with the
classification performance and a head-to-head comparison against the
ParT and SPANet baselines on the same data set, 
described in Section \ref{sec:event}.

\subsection{Motivation}
\label{sec:hgnn:motivation}
The high jet and lepton multiplicity of the $t\bar{t}t\bar{t}$ final
state, combined with the small signal-to-background ratio inherent to
the process, makes the analysis a natural target for modern deep
learning techniques. Conventional cut-based selections and low-dimensional
multivariate methods such as boosted decision trees operate on a
handful of high-level observables, but in $t\bar{t}t\bar{t}$ events the
relevant information is distributed across the full event: up to twelve
final-state partons accompanied by two to four prompt leptons, with
combinatorial ambiguity in their assignment to the four parent top
quarks. A model that consumes the full reconstructed event and learns
its own internal representation is therefore strongly preferred.

Among data-driven architectures, three large families have been
established as state of the art for jet- and event-level classification
at the LHC: convolutional networks acting on calorimeter
images~\cite{Kasieczka:2019dbj}, sequence-based architectures such as
the Particle Transformer and ParticleNet which treat the event as a
permutation-invariant point cloud of constituents~\cite{Qu:2022mxj,Qu:2019gqs},
and graph neural networks~\cite{Shlomi:2020gdn,Thais:2022iok} which encode pairwise
relations among reconstructed objects as graph edges. Graph based
approaches are particularly well matched to the inductive biases of
collider data: the input is permutation invariant, of variable
cardinality, and the relevant physics is local in $(\eta,\phi)$ and in
clustering distance.

Standard graph neural networks, however, only naively represent
\emph{pairwise} relations between reconstructed objects through edges.
The discriminating observables in $t\bar{t}t\bar{t}$ events are
intrinsically many-body: the invariant mass of a $b$-jet plus a charged
lepton plus the missing transverse momentum approximates the parent top
mass, the invariant mass of three jets reconstructs the hadronic top,
and the scalar sum of all jet $p_T$'s separates high-multiplicity
$t\bar{t}t\bar{t}$ events from $t\bar{t}V$ backgrounds. Encoding such
correlations in a graph that admits only pairwise edges requires
several rounds of message passing, in which information propagates
through intermediate nodes, and which therefore tends to dilute the
many-body signal across the network depth.

A \emph{hypergraph} generalizes a graph by allowing each edge, viz. termed
a hyperedge, to connect an arbitrary subset of nodes rather than just
two~\cite{Feng:2018hgnn,Bai:2021hgnnplus,pmlr-v139-zheng21b}. This is precisely the
representational primitive needed for many-body kinematic observables:
a single hyperedge can carry, as a learned feature, an invariant-mass
combination of any number of reconstructed objects, and the
corresponding message-passing operation propagates this combination to
all of its constituent nodes in a single layer. The remainder of this
section describes how we exploit this property for $t\bar{t}t\bar{t}$
identification.

\subsection{Event representation}
\label{sec:hgnn:representation}

Each reconstructed event in the combined multilepton signal region is
represented as a hypergraph $\mathcal{H} = (\mathcal{V}, \mathcal{E})$,
where $\mathcal{V}$ is the set of nodes and $\mathcal{E}$ the set of
hyperedges. Nodes correspond to the reconstructed final-state objects
that pass the selection criteria of
Section~\ref{sec:result}: isolated electrons and muons satisfying
the isolation requirement of Eq.~(\ref{eq:isolation}), and
anti-$k_T$~\cite{Cacciari:2008gp} jets with radius parameter $R = 0.4$,
transverse momentum $p_T > 25$~GeV, and pseudorapidity $|\eta| < 2.5$.
The missing transverse momentum vector $\vec{p}_T^{\,\mathrm{miss}}$ is
included as a dedicated global node, connected to every other node by
its participation in all hyperedges that contain a charged lepton.

\paragraph*{Node features.}
Each node $v \in \mathcal{V}$ carries a feature vector $h_v^{(0)} \in
\mathbb{R}^{d_{\mathrm{in}}}$ of dimension $d_{\mathrm{in}} = 13$, 
constructed from the following kinematic and identification
quantities:
\begin{enumerate}
\item four-momentum components $(p_T, \eta, \phi)$;
\item invariant mass of the jets;
\item jet radius in the pseudorapidity and azimuthal angle; 
\item ratio of the hadronic versus electromagnetic energy deposited in the calorimeter;
\item charged and neutral energy fraction of jets;
\item number of charged and neutral constituents in jet;
\item $b$-tagging discriminant score from the \textsc{Delphes} CMS-like tagger, set to zero for leptons.
\item energy weighted particle arrival time of flight in the calorimeter.
\end{enumerate}
For the dedicated $\vec{p}_T^{\,\mathrm{miss}}$ node, the rapidity and
energy entries are set to zero and the object-type encoding to a
fifth one-hot category. The resulting node features are standardized
to zero mean and unit variance over the training set.

\paragraph*{Hyperedge construction.}
We construct hyperedges of three orders. Pairwise hyperedges of order
two connect each pair of nodes within an angular distance
$\Delta R = \sqrt{\Delta\eta^2 + \Delta\phi^2} < 1.5$; 
they capture local correlations such as collinear splittings and
lepton--jet overlap. Three-body hyperedges of order three are
constructed by enumerating, for each $b$-tagged jet, all triplets
$(b, j_1, j_2)$ with $j_1, j_2$ light jets satisfying $\Delta R_{ij}$ threshold; 
these are designed to
expose the hadronic top kinematics $m_{bjj} \sim m_t$. 
Hyperedges with node cardinality four or higher
are constructed analogously, when decay product of top quarks are within geometric vicinity and fall below the hyperedge construction threshold. The total number of hyperedges varies event by event; in the combined
multilepton signal region we observe an average of
$\langle|\mathcal{E}|\rangle \approx 35$ 
per event, with order-two hyperedges dominating the count and order-four
hyperedges contributing the smallest fraction. 

\subsection{Network architecture}
\label{sec:hgnn:architecture}

The H-GNN tagger consists of a single message-passing block that
updates node, edge, and global-event representations over the
reconstructed \emph{pairwise} event graph, followed by a hyperedge
classification head that assigns scores to each 
hyperedge independently. A graph-level aggregation that pools the
resulting per-hyperedge predictions into one event-level score. Unlike
a stacked, recurrent message-passing scheme, a single such block is
applied once ($L=1$); unlike an attention-based hyperedge convolution,
the hypergraph structure enters only at the classification stage, not
in the message-passing step itself. A schematic of the architecture is
shown in Fig.~\ref{fig:hgnn_arch}.

\paragraph*{Message passing.}
There is no separate embedding stage: the raw per-object feature
vector $h_v^{(0)}$, the pairwise edge feature $e_{ij}^{(0)} = \Delta
R_{ij} \in \mathbb{R}$ defined for every pair with $\Delta R_{ij} <
1.5$, and the global feature $u^{(0)} = [\sum_v p_{T,v},\,\langle
\phi_v\rangle] \in \mathbb{R}^2$ are themselves the input to the first
layer of the update networks below. A single message-passing step
updates edges, then nodes, then the global feature, in turn,
\begin{align}
e'_{ij} &= \mathrm{MLP}_e\!\left(h_i^{(0)} \,\|\, h_j^{(0)} \,\|\, e_{ij}^{(0)} \,\|\, u^{(0)}\right),
\notag\\
m_{ij} &= \mathrm{MLP}_{n,1}\!\left(h_i^{(0)} \,\|\, h_j^{(0)} \,\|\, e'_{ij}\right),
\notag\\
h_j^{(1)} &= \mathrm{MLP}_{n,2}\!\Biggl(h_j^{(0)} \,\Big\|\, \underset{i\in\mathcal{N}(j)}{\mathrm{mean}}\, m_{ij} \,\Big\|\, \underset{i\in\mathcal{N}(j)}{\max}\, m_{ij} \,\Big\|\, u^{(0)}\Biggr),
\notag\\
u^{(1)} &= \mathrm{MLP}_g\!\left(u^{(0)} \,\Big\|\, \underset{v}{\mathrm{mean}}\, h_v^{(1)} \,\Big\|\, \underset{v}{\max}\, h_v^{(1)}\right),
\label{eq:hgnn_message_passing}
\end{align}
where $\mathrm{MLP}_e$, $\mathrm{MLP}_{n,1}$, $\mathrm{MLP}_{n,2}$, and
$\mathrm{MLP}_g$ are four independent three-layer perceptrons (ReLU~\cite{2018arXiv180308375A},
dropout $p=0.01$) with a common hidden width of $32$, and
$\mathcal{N}(j)$ is the set of nodes joined to $j$ by a pairwise edge.
The aggregation over $\mathcal{N}(j)$ in Eq.~(\ref{eq:hgnn_message_passing})
is a fixed mean-and-max pooling, with no learned attention weights and
no residual or normalization layers. We stress that this step
propagates information along the pairwise $\Delta R < 1.5$ graph, not
along the hyperedges of Section~\ref{sec:hgnn:representation}: the
hypergraph structure plays no role in the message-passing itself and
enters only in the classification step below.

\paragraph*{Hyperedge classification.}
The node and global representations $h_v^{(1)}$, $u^{(1)}$ produced by
Eq.~(\ref{eq:hgnn_message_passing}) are each passed through an
independent linear layer, $x_v^{t} = W_v h_v^{(1)} + b_v$ and $u^{t} =
W_g u^{(1)} + b_g$ ($\mathbb{R}^{32}\to\mathbb{R}^{32}$). Every
hyperedge 
is then scored independently by mean-pooling
the transformed features of its member nodes and concatenating the
result with the transformed global feature of its parent event,
\begin{align}
h_e &= \frac{1}{|e|}\sum_{v\in e} x_v^{t},
\notag\\
\ell_e &= \mathrm{MLP}_{\mathrm{cls}}\!\left(h_e \,\|\, u^{t}\right) \in \mathbb{R}^{K}\,.
\label{eq:hgnn_hyperedge_readout}
\end{align}
 $\mathrm{MLP}_{\mathrm{cls}}$ is a three-layer perceptron
($\mathbb{R}^{64}\!\to\!32\!\to\!16\!\to\!K$) with batch normalization,
ReLU, and dropout $p=0.01$ after each hidden layer. The
number of event classes, viz.  the $t\bar{t}t\bar{t}$ signal and the four
backgrounds $t\bar{t}W$, $t\bar{t}Z$, $t\bar{t}H$, $t\bar{t}VV$, denoted by $K~(=5)$,
as described in Section~\ref{sec:event}. The same $\mathrm{MLP}_{\mathrm{cls}}$
scores every hyperedge in an event.

\paragraph*{Readout and output.}
The event-level prediction is obtained by averaging the per-hyperedge
\emph{logits} -- not the pooled node representations -- over every
hyperedge $e$ belonging to the event $g$,
\begin{equation}
\ell_g = \underset{e \,\in\, g}{\mathrm{mean}}\; \ell_e \,\in\, \mathbb{R}^{K},
\qquad
\hat{p}(x) = \mathrm{softmax}\!\left(\ell_g\right) \,\in\, \Delta^{K-1}.
\end{equation}
That is, each hyperedge is classified first, by
Eq.~(\ref{eq:hgnn_hyperedge_readout}), and only the resulting
per-hyperedge class probabilities are subsequently pooled into a
single event-level prediction $\hat p(x)$, the reverse order to a
conventional pool-then-classify readout. The predicted class is
$\hat y = \arg\max_c \hat p_c(x)$, and the $t\bar t t\bar t$
discriminant used in Sections~\ref{sec:hgnn:training}--\ref{sec:result}
is the signal-class component $\hat p_{t\bar t t\bar t}(x)$.

The full network contains $18{,}869$ trainable parameters, making it
considerably smaller than the ParT baseline (${\sim}\,1.7\times 10^6$
parameters) and SPANet (${\sim}\,16.7 \times 10^6$) used as references
in Appendix~\ref{app:training_comparison}.

\subsection{Training procedure}
\label{sec:hgnn:training}
\begin{figure*}[!htb]
    \centering
    \includegraphics[width=0.49\textwidth, height=6cm]{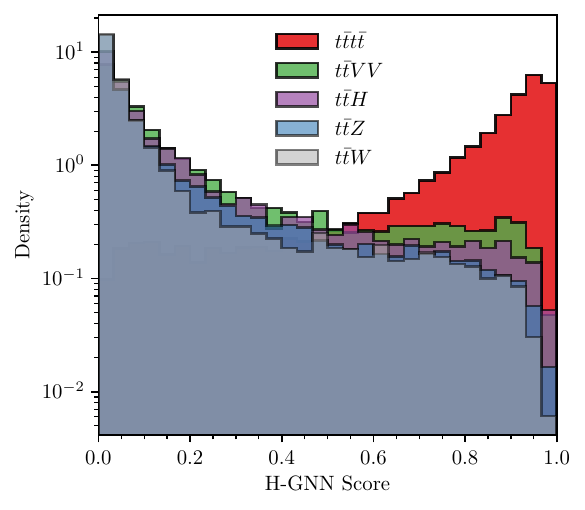}
    \includegraphics[width=0.49\textwidth, height=6cm]{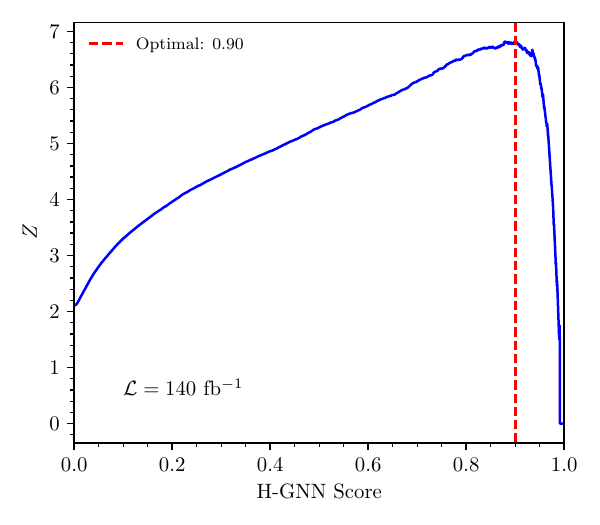}
    \caption{\label{fig:hgnn_score}Left: H-GNN score distributions for signal and background.
  Right: significance $Z$ versus H-GNN score threshold at $\mathcal{L}=140$~fb$^{-1}$, peaking at $0.90$ with $Z = 6.8\,\sigma$.}
\end{figure*}

The H-GNN tagger is trained as a multi-class classifier separating the $t\bar{t}t\bar{t}$ signal from four dominant irreducible backgrounds viz. $t\bar{t}W$, $t\bar{t}Z$, $t\bar{t}H$, and $t\bar{t}VV$ processes.
The training data set consists of $N_{\mathrm{train}} = 1.0 \times 10^6$ 
fully simulated events sampled in a $70{:}15{:}15$ train/validation/test
split, with class weights set to balance the five classes.

The loss function is the standard categorical cross-entropy
\begin{align}
    \mathcal{L}(g) = -\sum_{j=1}^5 y_j\,\mathrm{log}(p_j),
\end{align}
where $y$ is the one-hot and $g$ collectively denotes
the trainable network parameters. The minimization is performed with
the \textsc{AdamW} optimizer~\cite{adam} with an initial
learning rate of $3 \times 10^{-4}$ and a weight decay of $10^{-4}$,
using a cosine-annealing schedule over $50$ epochs and a batch size of
$256$. Gradients are clipped at a global norm of unity to prevent
training instabilities, and early stopping is triggered if the
validation loss does not improve over five consecutive epochs. Training is performed on a complete CPU environment (Intel Core i5-14500) and converges in
approximately seven hours.

Dropout with rate $p = 0.1$ is applied after each non-linear
transformation as a regularizer. The hyperparameters quoted above were
selected on the validation set by a small grid search over the
hypergraph convolution depth $L \in \{2, 3, 4\}$, hidden width
$d_h \in \{32, 64, 128\}$, and initial learning rate.

The H-GNN score distributions for signal and background are shown in Fig.~\ref{fig:hgnn_score} (left panel), with the discriminant separating the two classes most effectively in the high-score region. Prior to presenting the full set of WC constraints, we optimize the H-GNN discriminant threshold to maximize the statistical significance defined as~\cite{ParticleDataGroup:2024cfk}
\begin{align}
Z = \sqrt{2\,(N_S + N_B)\,\mathrm{ln}\Big(1 + \frac{N_S}{N_B}\Big) - N_S}.
\end{align}
The corresponding significance versus threshold distribution at an integrated luminosity $\mathcal{L}=140$~fb$^{-1}$ is shown in Fig.~\ref{fig:hgnn_score} (right panel), where the significance peaks at a threshold of $0.90$ with $Z = 6.8\,\sigma$. It shows a comparable significance to the $6.1\sigma$ reported by the ATLAS collaboration for a similar event selection.
\section{Analysis of SMEFT}
\label{sec:result}
\begin{figure}[!htb]
    \centering
    \includegraphics[width=0.98\linewidth]{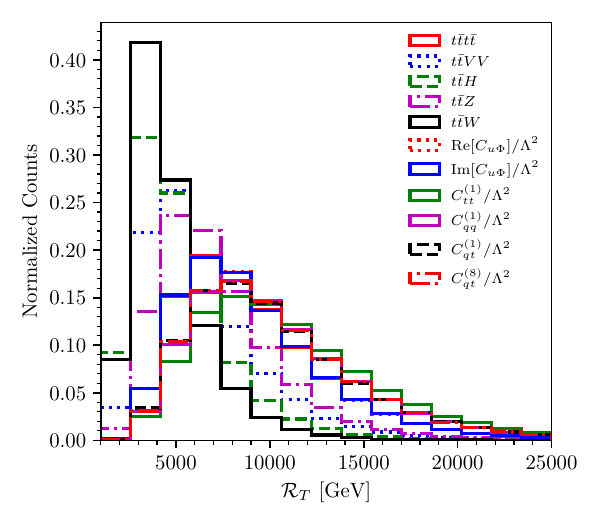}
    \caption{\label{fig:RT}Normalized distribution of the $\mathcal{R}_T$ observable for the $t\bar{t}t\bar{t}$ SM signal, dominant backgrounds ($t\bar{t}H$, $t\bar{t}Z$, $t\bar{t}W$), along with the sub dominant backgrounds like $t\bar{t}VV$, and representative benchmark values of the top-Yukawa and four-quark contact operators.}
\end{figure}

We perform a binned counting analysis to constrain the Wilson coefficients associated with the operators discussed in Section~\ref{sec:smeft}. Following event selection with the H-GNN tagger, we construct an event-wise single observable defined as
\begin{align}
    \label{eq:obs}
    \mathcal{R}_T = \sum_{a,b\in \mathrm{(Vis)}}\sqrt{p_T^{(a)}\,p_T^{(b)}\,\Delta R_{ab}} + \slashed{E}_T,
\end{align}
where $p_T^{(a,b)}$ is the transverse momenta of all visible particles (jets and leptons), $\Delta R_{ab}$ is the angular distance between the visible particles and $\slashed{E}_T$ denotes the missing transverse momenta. The corresponding normalized distribution of $\mathcal{R}_T$ is shown in Fig.~\ref{fig:RT}  for the $t\bar{t}t\bar{t}$ signal under the SM hypothesis and several WC benchmarks, together with the dominant background processes. The four-quark contact operators significantly enhance the high-$\mathcal{R}_T$ tail compared to the SM, while the top-Yukawa operator exhibits a suppression. This differential behavior demonstrates the sensitivity of $\mathcal{R}_T$ to the operator structure, with four-quark operators populating the tail through increased transverse activity, and the top-Yukawa operator shifting events toward lower $\mathcal{R}_T$ values. The events are binned into five disjoint regions of $\mathcal{R}_T$ as follows:
\begin{align}
    \mathcal{R}_T^{(n)} \in 
    \begin{cases}
        (0.0,\;5]~\mathrm{TeV}, & n=1,\\
        (5\,(n-1),\;5\,(n+1)]~\mathrm{TeV}, & n=2,\dots,4,\\
        (2~\mathrm{TeV},\;\infty), & n=5.
    \end{cases}
\end{align}

For a given  CP-even WC hypothesis $\{C_i\}$ viz. four contact interactions and real WC of $\mathscr{O}_{uH}$, the expected $\mathcal{R}_T$ weighted cross~section in bin $n$ is parametrized by the quadratic form
\begin{align}
    \sigma_n(C_i, C_j) = \sigma_n^{00} + \sum_{i} \sigma_n^{i}\, C_i + \sum_{i \ge j} \sigma_n^{ij}\, C_i C_j,
\end{align}
and for the CP-odd Im$[C_{uH}]$ WC, the weighted cross~section is parameterized as
\begin{align}
    \sigma_n(C_i) = \sigma_n^{00} + \sigma_n^i\,C_i^2,~C_i \equiv \mathrm{Im}[C_{u\Phi}].
\end{align}
Here, the coefficients $\sigma_n^{00}$, $\sigma_n^{i}$, and $\sigma_n^{ij}$ are extracted from Monte Carlo simulations reweighted by the appropriate NLO cross sections. For the $pp \to t\bar{t}t\bar{t}$ signal process at $\sqrt{13}$ TeV, we adopt a central value of $\sigma_{\text{NLO}} = 12~\text{fb}$ at NLO QCD+EW accuracy, corresponding to an NLO $k$-factor of $k_{t\bar{t}t\bar{t}} \approx 1.33$. For the $t\bar{t}H$, $t\bar{t}W$, and $t\bar{t}Z$ background processes, the NLO $K$-factors are taken from Ref.~\cite{Alwall:2014hca} as $k_{t\bar{t}H} = 1.28$, $k_{t\bar{t}Z}=1.4$, and $k_{t\bar{t}W}=1.5$. For $t\bar{t}VV$, we use $k_{t\bar{t}VV} = 1.36$~\cite{LHC:2016ypw}. The expected background yield in bin $n$ at integrated luminosity $\mathcal{L}$ receives contributions from the $t\bar{t}H$, $t\bar{t}Z$, $t\bar{t}W$, and $t\bar{t}VV$ processes as
\begin{align}
    B_n = \sigma_n^{00} + \sum_{i \,\in \,\mathrm{Bkgs.}} \frac{\sigma_i \cdot N_{n,i}^{\text{H-GNN}}}{N_{\text{MC}}} \times \mathcal{L},
\end{align}
where $N_{n,i}^{\text{H-GNN}}$ denotes the number of MC events passing the H-GNN selection in bin $n$ for process $i$, and $N_{\text{MC}}$ is the total number of generated events.
\begin{figure*}[!htb]
    \centering
    \includegraphics[width=0.49\textwidth]{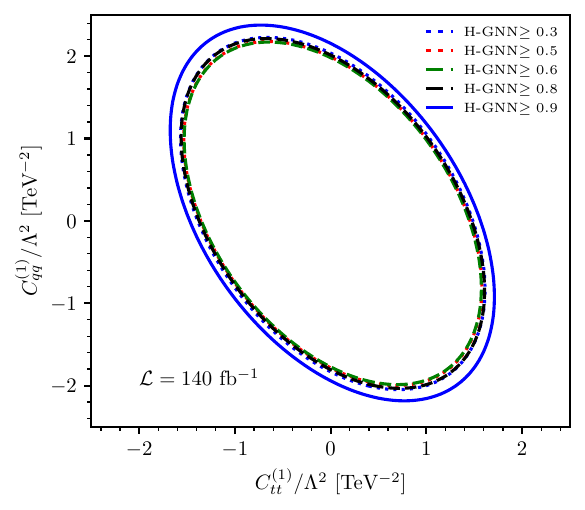}
    \includegraphics[width=0.49\textwidth]{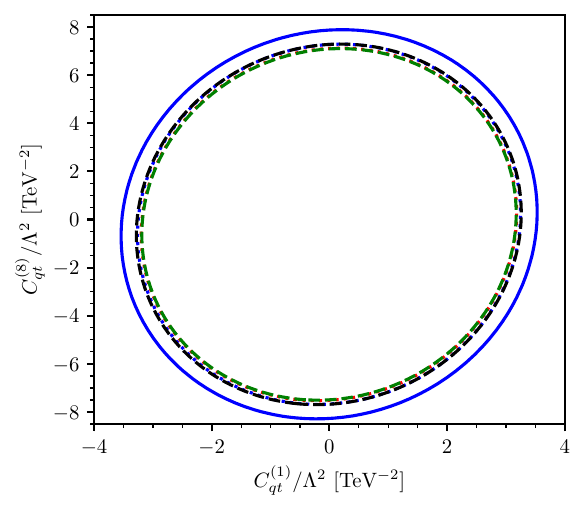}
    \caption{\label{fig:score} Two-dimensional $95\%$ CL contours for $\mathcal{R}_T$ observable in $C_{tt}^{(1)}-C_{qq}^{(1)}$ and $C_{qt}^{(1)}-C_{qt}^{(8)}$ planes for different values of threshold in Hyper-GNN output score at $\mathcal{L}=140$ fb$^{-1}$ with $50\%$ background systematic uncertainty.}
\end{figure*}
The sensitivity to the WCs is quantified via a $\chi^2$ test statistic
\begin{align}
    \Delta\chi^2 = \sum_{n=1}^5 \frac{\bigl[\sigma_n(C_i, C_j) - \sigma_n(0,0)\bigr]^2}{(\delta B_n)^2},
\end{align}
where the sum runs over all five bins of $\mathcal{R}_T$, and error on measurement on $\mathcal{R}_T$ is given by $$\delta B_n = \sqrt{B_n + \epsilon^2\, B_n^2}$$ where $\epsilon$ is the systematic uncertainty on the background yield in bin $n$. In the current work, we set the systematic uncertainty to a conservative value of $50\%$ following ATLAS~\cite{ATLAS:2023ajo} analysis.

Prior to presenting the full set of WC constraints, we also optimize the H-GNN discriminant threshold to maximize sensitivity to the WCs. To this end, we evaluate two-dimensional $95\%$ CL exclusion contours in the $(C_{tt}^{(1)},\, C_{qq}^{(1)})$ and $(C_{qt}^{(1)},\, C_{qt}^{(8)})$ planes at $\mathcal{L} = 140~\text{fb}^{-1}$ for a range of H-GNN output score thresholds (see Fig.~\ref{fig:hgnn_score} for score distribution). The resulting contours are shown in Fig.~\ref{fig:score}.

The $95\%$ CL contours in the $C_{tt}^{(1)}$--$C_{qq}^{(1)}$ plane exhibit only mild variation with the H-GNN score threshold. The contours for $0.3$ and $0.8$ are essentially identical, while those for $0.5$ and $0.6$ are slightly smaller. Although the significance is maximal at a threshold of $0.9$, the sensitivity is optimal at $0.6$, where the high-score tail carrying the largest SMEFT contribution is still retained. We consequently adopt a threshold of $0.6$ for the subsequent analyses, providing the best compromise between background suppression and retained signal statistics.

\begin{figure*}[!htb]
    \centering
    \includegraphics[width=0.98\linewidth]{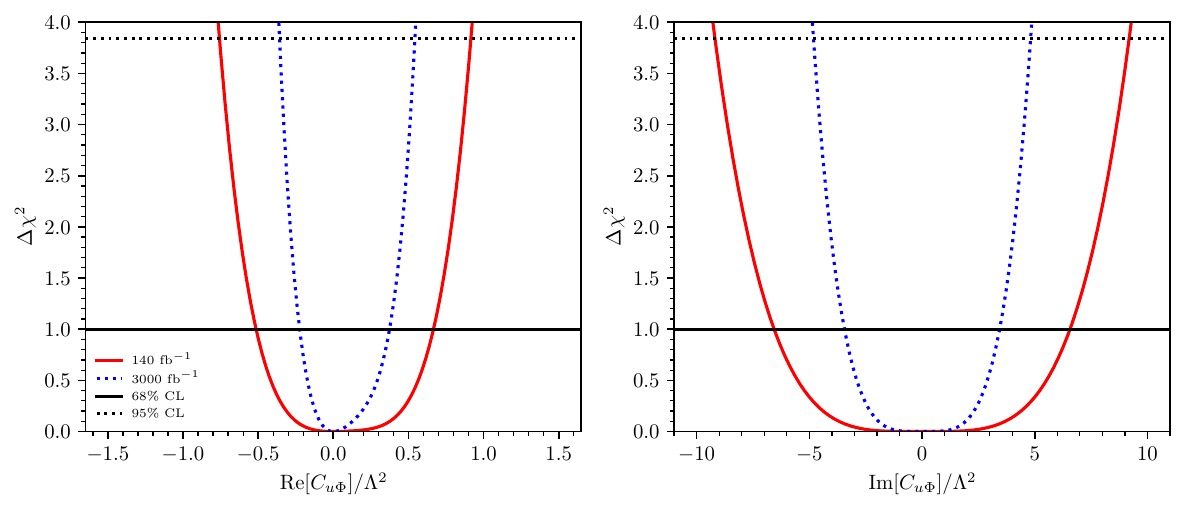}
    \caption{One dimensional distribution of $\Delta\chi^2$ as a function of real and imaginary part of $C_{u\Phi}$. The distribution are obtained at $\sqrt{s}=13$ TeV with an integrated luminosity $\mathcal{L}=140$ fb$^{-1}$ and $3000$ fb$^{-1}$ and systematic uncertainty of $50\%$ on backgrounds. The horizontal line at $\Delta\chi^2\sim 1.0$ corresponds to $68\%$ CL and at $3.84$ corresponds to $95\%$ CL.}
    \label{fig:ty1d}
\end{figure*}
Next, we analyze the $\mathscr{O}_{u\Phi}$, which governs the top-Yukawa coupling. The one-parameter $\Delta\chi^2$ profiles as a function of $\mathrm{Re}[C_{u\Phi}]$ and $\mathrm{Im}[C_{u \Phi}]$ are shown in Fig.~\ref{fig:ty1d} at $\mathcal{L} = 140~\text{fb}^{-1}$ (solid red curve) and $3000$ fb$^{-1}$ (dashed blue curve). The horizontal lines at $\Delta\chi^2 \sim 1.00$ and $3.84$ correspond to the $68\%$ and $95\%$ confidence level (CL) thresholds, respectively. The corresponding $95\%$ CL intervals at $140$ fb$^{-1}$ are
\begin{align}
    \mathrm{Re}[C_{u \Phi}]/\Lambda^2 &\in [-0.76,\;+0.92], \nonumber\\
    \mathrm{Im}[C_{u \Phi}]/\Lambda^2 &\in [-9.19,\;+9.19].
\end{align}
The wider bounds on imaginary part of $C_{u\Phi}$ compared to real part is due to the absence of interference term with the SM and the only contribution is coming at the quadratic level which is already $\mathcal{O}(1/\Lambda^4)$ suppressed. Inverting these bounds using Eq.~\eqref{eq:invert} into the anomalous coupling parametrization of Eq.~\eqref{eq:ltth} yields
\begin{align}
    a_t \in [+1.05,\,+0.94], \qquad |b_t| \lesssim 0.56.
\end{align}
The positive-definite interval on $a_t$ is consistent with the experimental constraint excluding negative values~\cite{ATLAS:2016neq}. Compared to the ATLAS~\cite{ATLAS:2023ajo} bound of $a_t \lesssim 2.2$ in the CP-even scenario, our constraint is tighter by approximately a factor of two. The result is also compatible with the CMS measurement $a_t \in [+0.86,\,+1.26]$~\cite{CMS:2022dbt}, obtained from multi-lepton $t\bar{t}H$ and $tH$ final states at $\sqrt{s} = 13~\text{TeV}$ and $\mathcal{L} = 138~\text{fb}^{-1}$. For the CP-odd parameter, our limit on $|b_t|$ improves upon the CMS bound of $|b_t| \lesssim 1.07$~\cite{CMS:2022dbt} by roughly a factor of $\sim 2$, demonstrating that the $t\bar{t}t\bar{t}$ process alone provides a competitive and complementary probe of CP violation in the top-Yukawa sector. The bounds on both real and imaginary WC of $\mathscr{O}_{\Phi u}$ further tightens by a factor of $\approx 2$ on increasing the luminosity to 3000 fb$^{-1}$.

\begin{figure*}[!htb]
    \centering
    \includegraphics[width=0.99\textwidth]{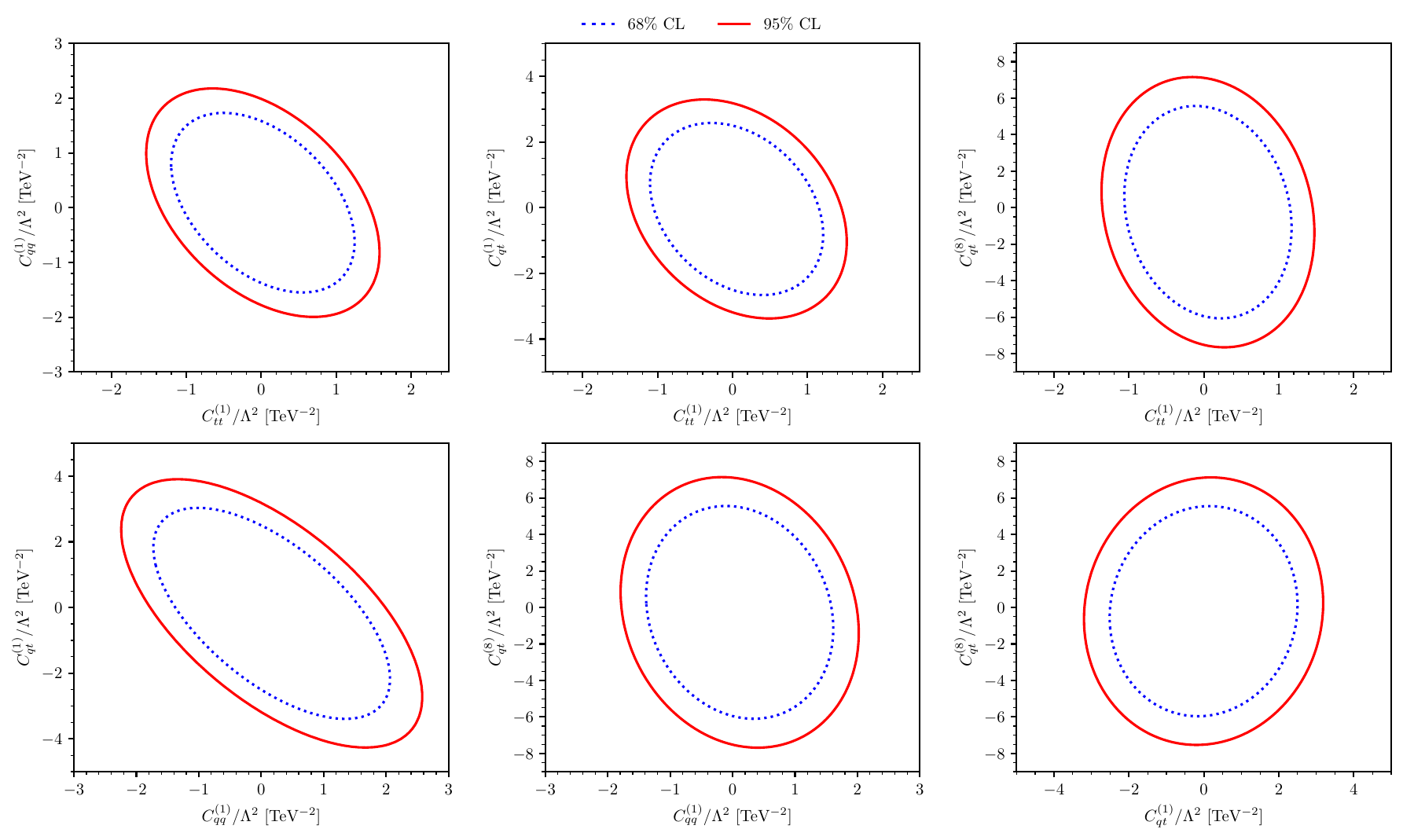}
    \caption{\label{fig:4ftwod}Likelihood $(\Delta\chi^2)$ distribution on varying two WCs at a time while keeping other two at zero with an integrated luminosity $\mathcal{L}=140$ fb$^{-1}$ and systematic error of $50\%$ on background estimation. The events are tagged with H-GNN score $\ge 0.6$. The two contours represented by solid blue and dashed blue corresponds to likelihood area at $68\%$ and $95\%$ CL, respectively.}
\end{figure*}

\begin{figure*}[!htb]
    \centering
    \includegraphics[width=0.495\textwidth]{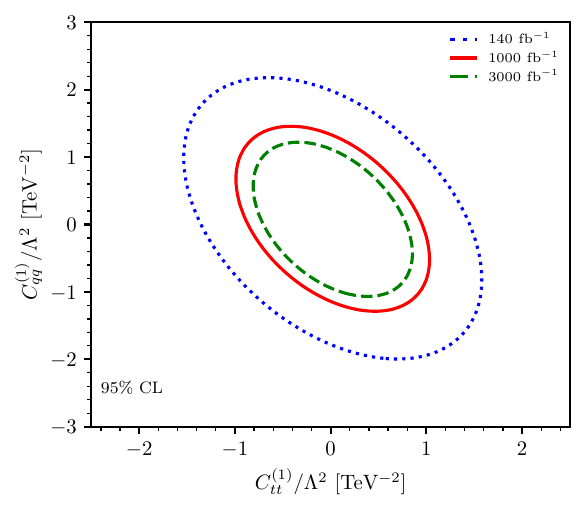}
    \includegraphics[width=0.495\textwidth]{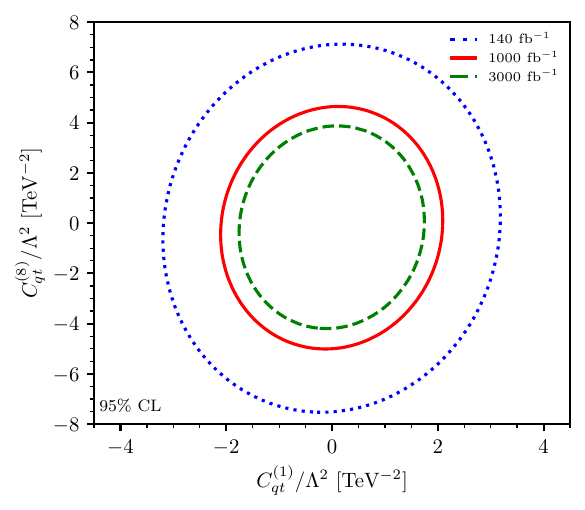}
    \caption{\label{fig:lumi}Two dimensional $95\%$ CL contours for $C_{tt}^{(1)}-C_{qq}^{(1)}$ (left panel) and $C_{qt}^{(1)} - C_{qt}^{(8)}$ (right panel) planes for three different values of integrated luminosity with systematic error of $50\%$ on background estimation. The events are tagged with H-GNN score $\ge 0.6$.}
\end{figure*}

\begin{table*}[!htb]
\centering
\caption{\label{tab:limitoneparadifflumi}
One-parameter limits on the Wilson coefficients ($C_i$ TeV$^{-2}$) of the four-fermion contact operators at $95\%$ confidence level using H-GNN tagger at an integrated luminosity of $140$ fb$^{-1}$ and systematic uncertainty of $50\%$ on the background estimation.
}
 \renewcommand{\arraystretch}{1.5}
 \begin{tabular*}{1\textwidth}{@{\extracolsep{\fill}}lccc@{}}\hline
  WCs & $140$ fb$^{-1}$ & $1000$ fb$^{-1}$ & $3000$ fb$^{-1}$ \\ \hline
  $C_{tt}^{(1)}/\Lambda^2$ & $[-1.21, +1.31]$ & $[-0.77, +0.86]$ & $[-0.63, +0.71]$ \\
  $C_{qq}^{(1)}/\Lambda^2$ & $[-1.58, +1.79]$ & $[-1.01, +1.20]$ & $[-0.83, +1.01]$ \\
  $C_{qt}^{(1)}/\Lambda^2$ & $[-2.85, +2.85]$ & $[-1.88, +1.87]$ & $[-1.57, +1.56]$ \\
  $C_{qt}^{(8)}/\Lambda^2$ & $[-6.75, +6.35]$ & $[-4.50, +4.13]$ & $[-3.77, +3.44]$ \\
\hline
\end{tabular*}
\end{table*}

We now present constraints on the WCs of the four-heavy-flavor fermion operators defined in Eq.~\eqref{eq:heavy}. Two-dimensional likelihood contours are computed by varying two WCs simultaneously while fixing the remaining two to zero, at an integrated luminosity of $\mathcal{L}=140$~fb$^{-1}$. The resulting $68\%$ and $95\%$ CL contours are shown in Fig.~\ref{fig:4ftwod} (solid red and dashed blue curves, respectively). Among the six possible coefficient pairs, three exhibit significant negative correlations: $(C_{tt}^{(1)},\,C_{qq}^{(1)})$, $(C_{tt}^{(1)},\,C_{qt}^{(1)})$, and $(C_{qq}^{(1)},\,C_{qt}^{(1)})$, while the remaining three pairs—$(C_{tt}^{(1)},\,C_{qt}^{(8)})$, $(C_{qq}^{(1)},\,C_{qt}^{(8)})$, and $(C_{qt}^{(1)},\,C_{qt}^{(8)})$—show minimal correlation, indicating orthogonal sensitivity from the chosen observables.

We further examine how these constraints evolve with increasing integrated luminosity, focusing on the HL-LHC scenario. Figure~\ref{fig:lumi} presents the $95\%$ CL contours for two representative planes: $C_{tt}^{(1)}-C_{qq}^{(1)}$ (left panel) and $C_{qt}^{(1)}-C_{qt}^{(8)}$ (right panel), at three benchmark luminosities: 140~fb$^{-1}$ (current LHC), $1000$~fb$^{-1}$ (HL-LHC baseline), and 3000~fb$^{-1}$ (HL-LHC upgrade). In both panels, the confidence ellipses shrink monotonically with increasing luminosity. However, the observed improvement from $140$~fb$^{-1}$ to $3000$~fb$^{-1}$ is approximately a factor of $\sim$2—significantly weaker than the $\sqrt{3000/140} \approx 4.6$ factor expected from pure statistical scaling—indicating that the sensitivity is systematics-dominated.

Table~\ref{tab:limitoneparadifflumi} summarizes the one-parameter $95\%$ CL intervals for all four WCs associated with four-quark contact operators at the three luminosity benchmarks with $50\%$ uncertainty in the background estimation. Increasing the integrated luminosity from $140$~fb$^{-1}$ to $1000$~fb$^{-1}$ improves the bounds by a factor of approximately $1.64$, while the further increase to $3000$~fb$^{-1}$ yields an additional improvement factor of $1.3$. The overall improvement from $140$~fb$^{-1}$ to $3000$~fb$^{-1}$ is thus about a factor of $\sim 2.2$ for all WCs. Comparing with the ATLAS limits at $\mathcal{L}=140$~fb$^{-1}$ listed in Table~\ref{tab:atlassmeft}, our constraints show varying levels of improvement depending on the operator. For $C_{qt}^{(1)}$ and $C_{qt}^{(8)}$, our limits are similar to that of the ATLAS (see Table~\ref{tab:atlassmeft}). Notably, our constraints on $C_{qq}^{(1)}$ are significantly stronger by a factor of $\sim 2.3$. For $C_{tt}^{(1)}$, our limits are slightly tighter by a factor of $\sim 1.5$.

\begin{figure*}[!htb]
    \centering
    \includegraphics[width=0.99\textwidth]{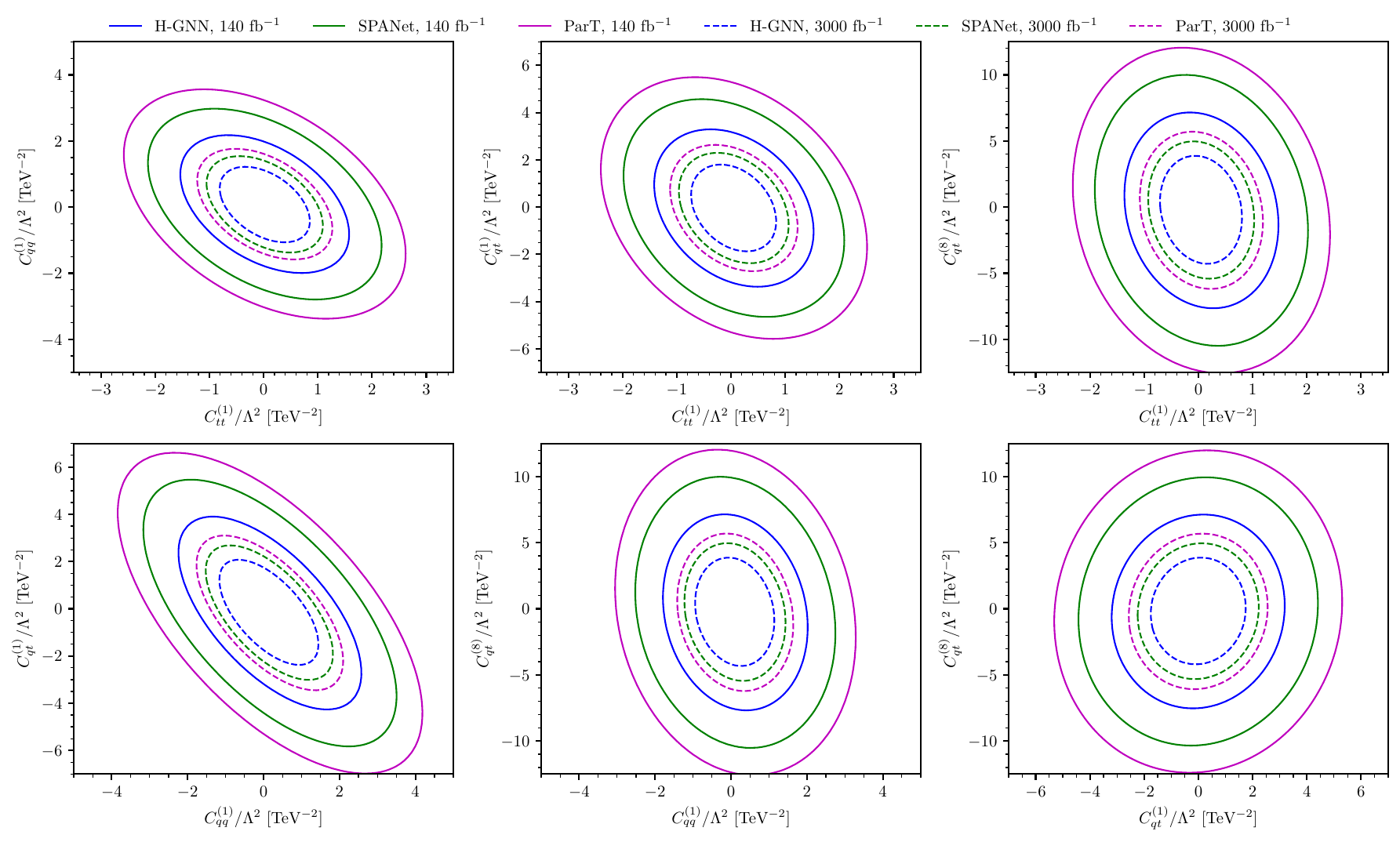}
    \caption{\label{fig:2dml}Two dimensional $\Delta\chi^2$ contours at $95\%$ CL for all six pairs of contact operators obtained from three different tagger at two values of luminosity viz. $\mathcal{L}=140$ fb$^{-1}$ (solid curves) and $\mathcal{L}=3000$ fb$^{-1}$ (dashed curves).}
\end{figure*}

Finally, we make an explicit comparison on the bounds on four-fermion contact operators based on event-tagger with three different neural networks viz., H-GNN, SPANet and ParT at threshold score of $0.6$. For that, we compute two dimensional likelihood contours for all six combinations of WCs using all three taggers, see Fig.~\ref{fig:2dml}. The $95\%$ CL contours are drawn at two values of luminosity viz., 140 $^{-1}$ (solid curve) and 3000 fb$^{-1}$ (dashed curve) with $50\%$ systematic uncertainty on background. From the figure, it becomes evident that the tightest bounds on all WCs are provided by H-GNN followed by SPANet and ParT. It clearly highlights the effectiveness of using H-GNN on increasing the signal-background ratio in the case of $t\bar{t}t\bar{t}$ process at LHC.

\section{Discussion and Conclusion}
\label{sec:conclusion}

We have presented a phenomenological study of four-top-quark production,
$pp \to t\bar{t}t\bar{t}$, at the LHC at $\sqrt{s} = 13$~TeV, employing
a Hyper-Graph Neural Network (H-GNN) as the primary event classifier.
Combining same-sign dilepton, trilepton, and four-lepton signal regions
following a CMS-like event selection, and discriminating the
$t\bar{t}t\bar{t}$ signal from the dominant $t\bar{t}V$, $t\bar{t}H$,
and multi-boson backgrounds, the H-GNN achieves an area under the ROC
curve of $\mathrm{AUC} = 0.97$ and an optimal statistical significance to be $6.8$ at an integrated luminosity of
$\mathcal{L} = 140~\mathrm{fb}^{-1}$ with a classifier threshold at $0.9$. This represents a substantial improvement over the SPANet $Z = 3.5$ and ParT $Z=2.6$ baselines trained and evaluated on the same
dataset~\cite{Shmakov:2021qdz,Qu:2022mxj}, and a comparable optimal significance reach to that of the ATLAS $Z=6.1\sigma$~\cite{ATLAS:2023ajo} under identical event selection criteria.

The performance advantage of the H-GNN over conventional graph and
sequence-based architectures stems from its hypergraph representation of
the collision event, in which hyperedges of order two through four encode
pairwise, hadronic-top ($m_{bjj}$), and semileptonic-top ($m_{b\ell\nu}$)
kinematic combinations directly at the architectural level. These many-body
correlations, which are the primary discriminants against $t\bar{t}V$ and
$t\bar{t}H$ backgrounds in the multilepton final state, would otherwise
require deep iterated message passing in a standard graph architecture to
emerge. Notably, the H-GNN achieves this discriminating power with only
${\sim}\,1.9 \times 10^4$ trainable parameters, rendering it considerably
more compact than the ParT (${\sim}\,2.0 \times 10^6$)
and SPANet (${\sim}\,16.7 \times 10^6$) baselines, and providing a
resource-efficient alternative for high-multiplicity final-state analyses.

We exploited the improved signal extraction to perform an SMEFT analysis
constraining the Wilson coefficients of five dimension-six operators that
affect the $t\bar{t}t\bar{t}$ rate: the top-Yukawa operator
$\mathscr{O}_{u\Phi}$~\cite{Grzadkowski:2010es} and the four-heavy-fermion contact operators
$\mathscr{O}_{tt}^{(1)}$, $\mathscr{O}_{qq}^{(1)}$, $\mathscr{O}_{qt}^{(1)}$,
and $\mathscr{O}_{qt}^{(8)}$~\cite{DHondt:2018cww}. For the top-Yukawa
sector, we obtain the one-parameter $95\%$ CL intervals
$\mathrm{Re}[C_{u \Phi}] \in [-0.76,\,+0.92]$ and
$\mathrm{Im}[C_{u \Phi}] \in [-9.19,\,+9.19]$, which translate in the
anomalous coupling parameterization of Eq.~\eqref{eq:ltth} to
$a_t \in [+1.05,\,+0.94]$ and $|b_t| \lesssim 0.56$. The constraint on the
CP-even parameter $a_t$ is tighter than the ATLAS bound
$|a_t| < 2.2$~\cite{ATLAS:2023ajo} by approximately a factor of two,
while the constraint on the CP-odd parameter $|b_t|$ improves upon the
CMS bound obtained from $t\bar{t}H$ and $tH$ multilepton
analyses~\cite{CMS:2022dbt} by a factor of $\sim 2$, demonstrating the unique
sensitivity of the $t\bar{t}t\bar{t}$ process to CP-violating top-Yukawa
interactions. For the four-heavy-fermion operators,
the one-parameter $95\%$ CL bounds at $\mathcal{L} = 140~\mathrm{fb}^{-1}$
tighten the ATLAS limits~\cite{ATLAS:2023ajo} by a factor of ${\sim}\,1.5$ and $\sim 2.3$
for $C_{tt}^{(1)}$ and $C_{qq}^{(1)}$, respectively. Our limits for $C_{qt}^{(1)}$ and $C_{qt}^{(8)}$ remains similar to that of the ATLAS. The two-dimensional likelihood contours
(Fig.~\ref{fig:4ftwod}) reveal significant anti-correlations between the
pairs $(C_{tt}^{(1)},\,C_{qq}^{(1)})$, $(C_{tt}^{(1)},\,C_{qt}^{(1)})$,
and $(C_{qq}^{(1)},\,C_{qt}^{(1)})$, while the pairs involving
$C_{qt}^{(8)}$ remain largely uncorrelated, indicating orthogonal
sensitivity between the octet and singlet contact interactions.

Projecting to the HL-LHC, we find that the confidence intervals contract
by a factor of approximately $1.5$ - $1.7$ at $1000~\mathrm{fb}^{-1}$ and
by a factor of ${\sim}\,2.0$ - $2.1$ at $3000~\mathrm{fb}^{-1}$ relative
to $140~\mathrm{fb}^{-1}$, for all four-fermion operator coefficients.
The lower than expected statistical scaling of these improvements, compared to the
$\sqrt{3000/140} \approx 4.6$ factor expected from pure statistical
scaling—indicates that the analysis is entering a systematics-limited
regime at HL-LHC luminosities, motivating future efforts to reduce
theoretical and experimental uncertainties on the dominant $t\bar{t}W$,
$t\bar{t}Z$, $t\bar{t}H$, and $t\bar{t}VV$ background processes. 

We also compared the sensitivity of the three event taggers, namely H-GNN, SPANet, and ParT, by deriving two-dimensional likelihood contours for all six combinations of four-fermion contact operators. The $95\%$ CL limits were obtained for integrated luminosities of $140~\mathrm{fb}^{-1}$ and $3000~\mathrm{fb}^{-1}$, assuming a $50\%$ systematic uncertainty on the background. Among the three approaches, H-GNN consistently provides the strongest constraints on all WCs, followed by SPANet and ParT.

Several directions remain open for future work. The present analysis
adopts a simplified treatment of systematic uncertainties; a full
profile-likelihood fit incorporating correlated shape and rate
uncertainties for each background process, together with $b$-tagging
efficiency and jet energy scale uncertainties, would be required for
a direct comparison with experimental results and to assess the
robustness of the derived SMEFT bounds under realistic detector
conditions. The multilepton channel studied here could be complemented
by an analysis of the same-sign dilepton channel with resolved
top-quark reconstruction, which provides additional sensitivity to
the high-$H_T$ tails where the four-fermion operators have the
largest impact. Furthermore, the hypergraph architecture introduced here
is not specific to $t\bar{t}t\bar{t}$ production and can be extended
to other rare multi-top signatures accessible at the LHC, such as
$t\bar{t}tW$, $t\bar{t}tZ$, and associated three-top processes, as
well as to broader high-multiplicity BSM searches. Finally, at
future hadron colliders operating at $\sqrt{s} = 100$~TeV, such as
the FCC-hh~\cite{Benedikt:2022kan,FCC:2018vvp}, the four-top cross
section increases by nearly two orders of magnitude, greatly enhancing
the statistical reach for SMEFT studies; the H-GNN framework
developed here provides a natural starting point for such analyses.

In summary, we have demonstrated that hypergraph-based neural network
architectures provide a powerful and physically motivated tool for the
classification and SMEFT interpretation of rare multi top processes at
the LHC. The H-GNN outperforms state-of-the-art graph and
attention-based baselines while remaining computationally efficient,
and the resulting SMEFT bounds on top-quark interactions represent a
factor of two to four improvement over existing experimental limits.
These results establish hypergraph networks as a valuable addition to
the toolkit for precision top-quark phenomenology at present and future
hadron colliders.

\section{Acknowledgments}
\label{sec:acknowledgment}
\noindent
 S.G. is supported by the IIT-Kanpur faculty initiation grant (IITK /PHY /2023499) and Anusandhan National Research Foundation, Advanced Research Grant (ANRF/ARG/2025/005801/PS) and Anusandhan National Research Foundation -
ANRF MISSION AI for Science and Engineering Grant(ANRF/AISE/2025/000373/AISSAI. \\
The authors would like to thank the ICTS program : Statistical Methods and Machine Learning in High Energy Physics (https://www.icts.res.in/program/ML4HEP), from where the project was conceptualized. 
\appendix

\section*{Data and Code Availability}
The data that support the findings of this study are openly available in Ref.~\cite{subba_2026_24101367}. The analysis code used to generate the results reported here is available at \url{https://github.com/amirsubba/Four-Top-quarks-H-GNN_2605.18382}.
\begin{figure}[!tp]
 \centering
  \hspace{-0.5cm}
  \includegraphics[height=0.9\textheight, keepaspectratio]{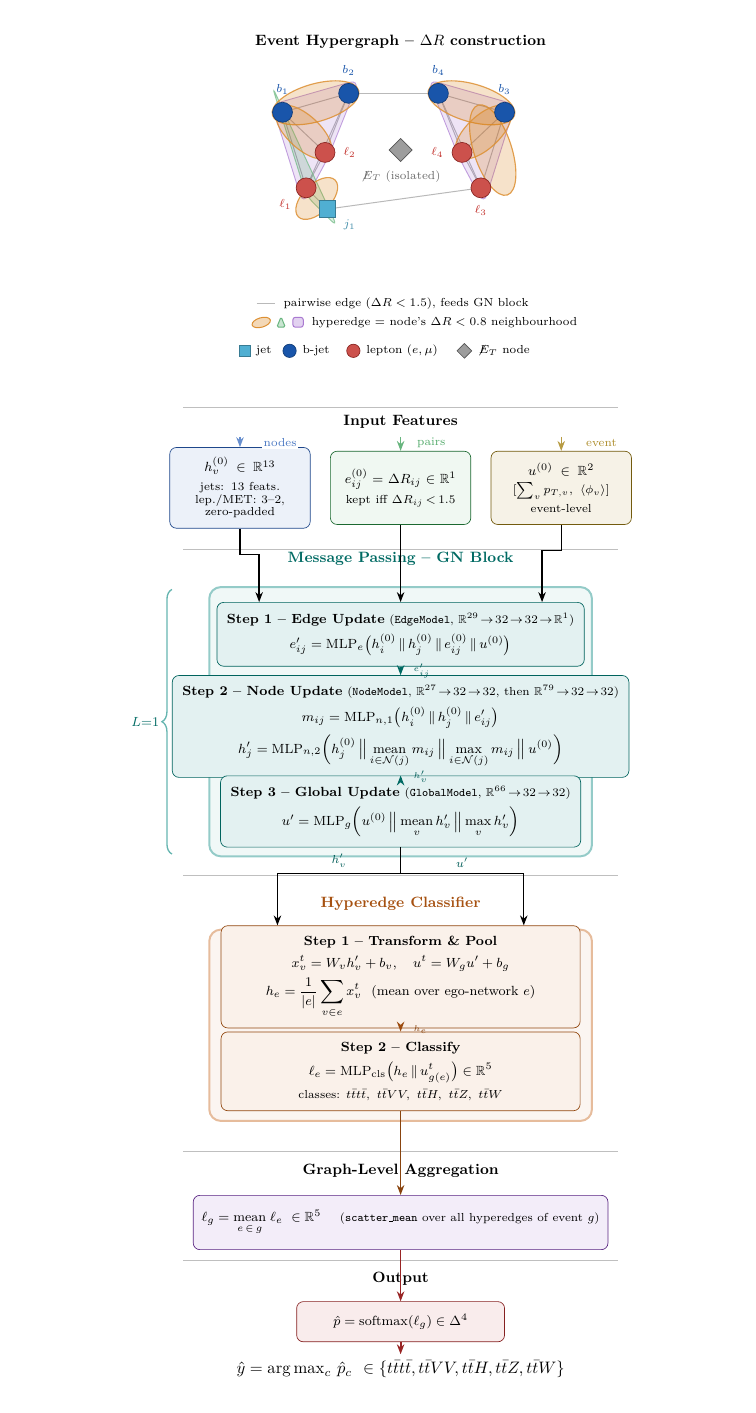} 
  \caption{Schematic of the Hyper-Graph Neural Network (H-GNN)
    classifier used in this work, illustrated on a representative
    SR-$4\ell$ event topology with four prompt leptons
    ($\ell_{1,\ldots,4}$), four $b$-tagged jets ($b_{1,\ldots,4}$),
    and the missing transverse-momentum node ($\protect\slashed{E}_T$). A single light jet $j_1$ has been kept in the event as a representative of the fact that in general there can be light jets in the final state.
    }
  \label{fig:hgnn_arch}
\end{figure}

\section{Baseline Network Architectures}
\label{app:networks}

This appendix provides self-contained descriptions of the two baseline
classifiers—the Particle Transformer (ParT) and SPANet—against which the
H-GNN is benchmarked in Section~\ref{sec:hgnn}.  Both baselines receive
the same per-object input features as the H-GNN node features.
The basic construction of the H-GNN architecture is described in Section~\ref{sec:hgnn:representation} and displayed in Fig.~\ref{fig:hgnn_arch}.
The baseline models are trained under identical
conditions (Section~\ref{sec:hgnn:training}), so that observed performance
differences are attributable solely to the inductive biases of each
architecture.

\subsection{Particle Transformer}
\label{app:part_transformer}

The Particle Transformer (ParT)~\cite{Qu:2022mxj} adapts the standard
Transformer encoder~\cite{Vaswani:2017attention} to the point-cloud
representation of a collider event, augmenting the self-attention mechanism
with learned pairwise interaction terms that capture the geometry of the
final state.

\paragraph*{Input representation.}
Each event is treated as an unordered set of $N$ objects
(jets, leptons, and the $\vec{p}_T^{\,\mathrm{miss}}$ node).  Object $i$
carries the same 12-dimensional feature vector
$\mathbf{x}_i\in\mathbb{R}^{12}$ used as H-GNN node features.  In
addition, ParT constructs a six-dimensional pairwise interaction feature
for every ordered pair $(i,j)$,
\begin{equation}
  u_{ij} = \bigl(\Delta\eta_{ij},\;\Delta\phi_{ij},\;
                 \ln\Delta R_{ij},\;
                 \ln k_{T,ij},\;\ln z_{ij},\;
                 \ln m_{ij}\bigr),
  \label{eq:part_pair}
\end{equation}
where
\begin{align}
  k_{T,ij} &= \min(p_{T,i},\,p_{T,j})\,\Delta R_{ij}/R_0, \\
  z_{ij}   &= \min(p_{T,i},\,p_{T,j})\,/\,(p_{T,i}+p_{T,j}), \\
  m_{ij}   &= \text{invariant mass of the $(i,j)$ system},
\end{align}
with $R_0=0.4$.  These quantities encode the collinear and soft structure
of the pair and supply information that is physically motivated by
QCD radiation patterns.

\paragraph*{Interaction-augmented self-attention.}
The interaction features $u_{ij}$ are embedded into a per-head attention
bias via a two-layer MLP,
$U_{ij}^{(h)} = \mathrm{MLP}_U(u_{ij})\in\mathbb{R}$, and added to the
standard scaled dot-product attention score:
\begin{align}
  A^{(h)}_{ij}
  &= \frac{\mathbf{q}^{(h)}_i \cdot \mathbf{k}^{(h)}_j}{\sqrt{d_k}}
  + U^{(h)}_{ij},\nonumber\\
  \mathbf{o}^{(h)}_i
  &= \sum_j \mathrm{softmax}\bigl(A^{(h)}_{\cdot j}\bigr)_i\;
           \mathbf{v}^{(h)}_j,
  \label{eq:part_attn}
\end{align}
where $\mathbf{q}^{(h)}_i, \mathbf{k}^{(h)}_j, \mathbf{v}^{(h)}_j
\in\mathbb{R}^{d_k}$ are the query, key, and value projections for head
$h$.  Multi-head outputs are concatenated and projected back to the model
dimension $d_h$, after which a position-wise feed-forward network (FFN)
and LayerNorm residual connections~\cite{Ba:2016layernorm} complete
each encoder block.

\paragraph*{Class-attention readout.}
After $L_\mathrm{ParT}$ standard encoder blocks, ParT appends a learnable
class token $\mathbf{c}^{(0)}\in\mathbb{R}^{d_h}$ and runs $L_\mathrm{CA}$
\emph{class-attention} blocks in which the class token attends to all
particle tokens but the particle tokens do not attend to each other:
\begin{equation}
  \mathbf{c}^{(\ell+1)} = \mathrm{LayerNorm}\!\Bigl(
    \mathbf{c}^{(\ell)} +
    \mathrm{MHA}\!\bigl(\mathbf{c}^{(\ell)},\,
                         \{\mathbf{h}_i^{(L_\mathrm{ParT})}\}\bigr)
  \Bigr),
\end{equation}
where $\mathrm{MHA}(\cdot,\cdot)$ denotes multi-head cross-attention.
The final class token $\mathbf{c}^{(L_\mathrm{CA})}$ is passed through a
two-layer MLP with sigmoid activation to produce the binary
$t\bar{t}t\bar{t}$ classifier score.  In our implementation, the
hyperparameters $L_\mathrm{ParT}=8$, $L_\mathrm{CA}=2$, $d_h=128$, and
$n_\mathrm{heads}=8$ are adopted from the original jet-tagging
configuration, yielding approximately $2.0\times 10^6$ trainable
parameters.

\subsection{SPANet}
\label{app:spanet}

SPANet~\cite{Shmakov:2021qdz} (Symmetry-Preserving Attention Network) was
originally developed for the combinatorial jet-assignment problem in
top-quark pair reconstruction, where the goal is to map a set of
reconstructed jets onto the decay products of each top quark while
respecting the permutation symmetries of the assignment.  We re-purpose its
encoder as a permutation-invariant event-level classifier for the
$t\bar{t}t\bar{t}$ signal versus background task.

\begin{figure*}[!htb]
    \centering
    \includegraphics[width=0.49\textwidth]{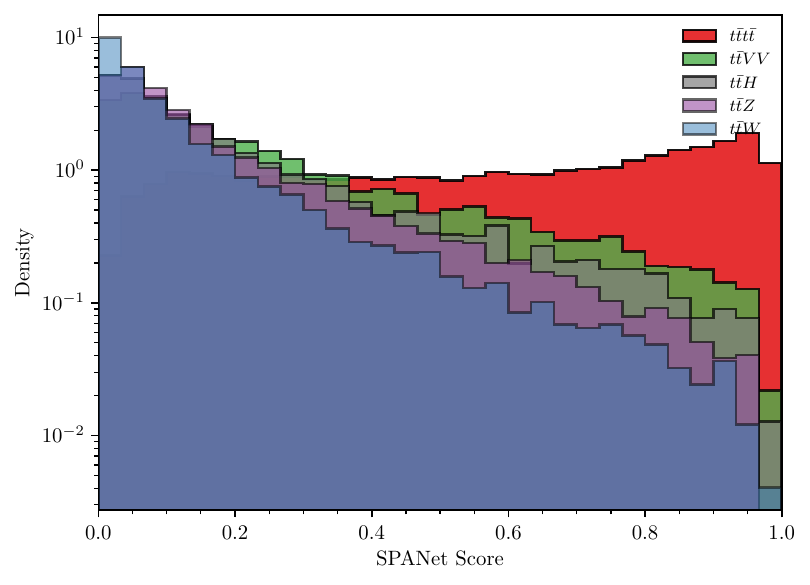}
    \includegraphics[width=0.49\textwidth]{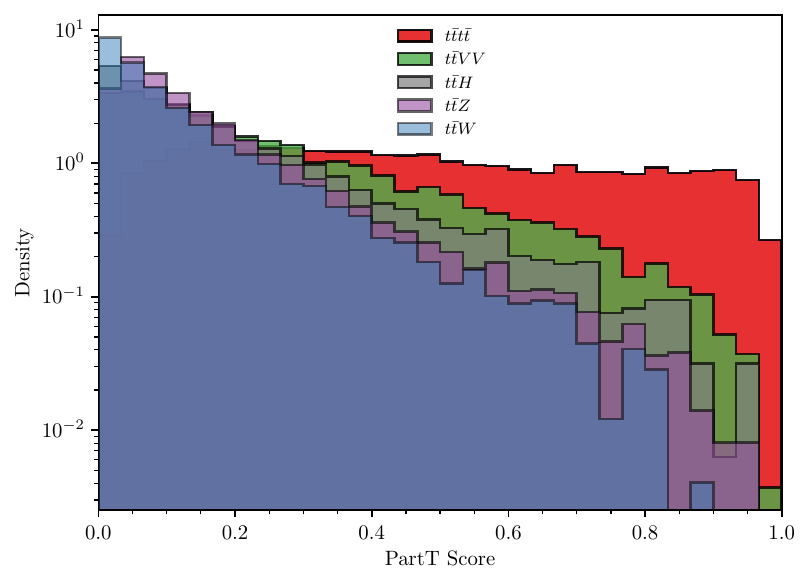}
    \caption{Comparison of the output scores for the two different architectures, viz. SPANet \& ParT
    for the signal and background samples (see Fig.~\ref{fig:hgnn_score} for H-GNN score).}
    \label{fig:model_score}
\end{figure*}

\paragraph*{Input embedding.}
Each reconstructed object $i$ is linearly embedded into a
$d_\mathrm{SPA}$-dimensional hidden vector,
\begin{equation}
  \mathbf{h}_i^{(0)} = W_{\mathrm{in}}\,\mathbf{x}_i + \mathbf{b},
  \qquad \mathbf{h}_i^{(0)}\in\mathbb{R}^{d_\mathrm{SPA}}.
\end{equation}

\paragraph*{Symmetry-preserving attention encoder.}
The central component of SPANet is a stack of $L_\mathrm{SPA}$
transformer-like encoder layers augmented with a
\emph{symmetry-preserving} attention mechanism.  The attention kernel is
designed to respect the permutation symmetries of the $t\bar{t}t\bar{t}$
final state: jets that are interchangeable by some symmetry of the process
(e.g., the two light-quark jets from a hadronic $W$ decay) contribute
symmetrically to the event representation.  At each layer $\ell$,
multi-head self-attention is applied over the full set of objects:
\begin{equation}
  \tilde{\mathbf{h}}_i^{(\ell)}
  = \mathrm{LayerNorm}\!\Bigl(
    \mathbf{h}_i^{(\ell)} +
    \sum_{j} a^{(\ell)}_{ij}\,V^{(\ell)}\mathbf{h}_j^{(\ell)}
  \Bigr),
\end{equation}
\begin{equation}
  \mathbf{h}_i^{(\ell+1)}
  = \mathrm{LayerNorm}\!\Bigl(
    \tilde{\mathbf{h}}_i^{(\ell)} +
    \mathrm{FFN}\!\bigl(\tilde{\mathbf{h}}_i^{(\ell)}\bigr)
  \Bigr),
\end{equation}
where the attention weights $a^{(\ell)}_{ij}$ are computed with
softmax-normalized dot-product attention and the permutation structure is
enforced by partitioning objects into symmetry groups prior to the
attention computation.

\paragraph*{Event-level classification.}
For the binary classification task considered here, we extract a global
event representation by mean-pooling the final-layer hidden states over
all $N$ objects,
\begin{equation}
  \mathbf{h}_g = \frac{1}{N}\sum_{i=1}^{N}\mathbf{h}_i^{(L_\mathrm{SPA})},
\end{equation}
and pass $\mathbf{h}_g$ through a two-layer MLP with sigmoid activation to
produce the $t\bar{t}t\bar{t}$ score.  We use $L_\mathrm{SPA}=6$,
$d_\mathrm{SPA}=64$, $n_\mathrm{heads}=4$, giving approximately
$5.0\times 10^5$ trainable parameters.


\subsection{Common training configuration and performance summary}
\label{app:training_comparison}

Both baselines are trained under conditions identical to those described
for the H-GNN in Section~\ref{sec:hgnn:training}: the same
$70$:$15$:$15$ train/validation/test event split, the AdamW
optimizer~\cite{Loshchilov:2017adamw} with initial learning rate
$3\times 10^{-4}$ and weight decay $10^{-4}$, cosine-annealing over
$50$ epochs, binary cross-entropy loss, and early stopping on the
validation loss.  Dropout at rate $p=0.1$ is applied throughout.
Training is performed on the same Intel Core i5-14500 used for the H-GNN.

Fig.~\ref{fig:hgnn_score} and Fig.~\ref{fig:model_score} presents the output scores by the three different architectures on the signal and background samples. We find that the statistical discrimination power of H-GNN outperforms the other architectures. Table~\ref{tab:arch_comparison} summarizes the three architectures and
their classification performance on the combined multilepton signal
region at $\mathcal{L}=140~\mathrm{fb}^{-1}$.  The H-GNN achieves the
highest significance despite having the fewest parameters, a consequence
of the explicit encoding of many-body kinematic correlations through
hyperedges of order three and four that represent hadronic- and
semileptonic-top combinations directly at the architectural level.  The
ParT and SPANet baselines, which rely on pairwise or point-cloud
representations, must recover these many-body structures implicitly
through multiple rounds of message-passing or attention, resulting in
lower discriminating power for the $t\bar{t}t\bar{t}$ task.

\begin{table}[!htb]
  \centering
  \renewcommand{\arraystretch}{1.4}
  \caption{\label{tab:arch_comparison}
  Summary of the three network architectures benchmarked in this work. The AUC is evaluated on the $t\bar{t}t\bar{t}$ vs. combined background ROC curve on the held-out test set.  The statistical significance $Z$ is computed at $\mathcal{L}=140~\mathrm{fb}^{-1}$ under the event selection of Section~\ref{sec:event}.  The ATLAS analysis value $Z=6.1$ is included for reference under the same event selection criteria.}
  \begin{tabular*}{0.48\textwidth}{@{\extracolsep{\fill}}lcccc@{}}
    \hline
    Architecture & Input repr.\ & Params & AUC & $Z$\;(140\,fb$^{-1}$)  \\
    \hline
    H-GNN (this work)  & hypergraph  & $1.9\times10^4$ & $0.97$ & $6.8$ \\
    SPANet~\cite{Shmakov:2021qdz}
                       & point cloud & $16.7\times10^6$ & $0.87$     & $3.5$ \\
    ParT~\cite{Qu:2022mxj}
                       & point cloud & $1.72\times10^6$ & $0.83$     & $2.6$ \\
    ATLAS~\cite{ATLAS:2023ajo}
                       & Graph         & ---             & ---     & $6.1$ \\
    \hline
  \end{tabular*}
\end{table}

\bibliography{refer}

\end{document}